\documentclass[fleqn,usenatbib]{mnras}

\usepackage{newtxtext,newtxmath}

\usepackage[T1]{fontenc}
\usepackage{ae,aecompl}

\usepackage{graphicx}	
\usepackage{amsmath}	
\usepackage{amssymb}	
\usepackage{lscape}
\usepackage{longtable}

\usepackage{amssymb,xcolor}
\usepackage{colortbl}

\newcommand{\swift}{{\it Swift}}
\newcommand{\rxte}{{\it RXTE}}
\newcommand{\integral}{{\it INTEGRAL}}
\newcommand{\grs}{GRS 1716--249}
\newcommand{\gro}{GRO J1719--24}
\newcommand{\swf}{Swift J1745--26}
\newcommand{\cnt}{ct/s}
\newcommand{\cm}{cm$^{-2}$}
\newcommand{\ergs}{erg s$^{-1}$}

\newcommand{\ergcm}{erg cm$^{-2}$ s$^{-1}$}
\newcommand{\kti}{kT$_{\rm in}$}
\newcommand{\kte}{kT$_{\rm e}$}
\newcommand{\chir}{$\chi^{2}_{\nu}$}
\newcommand{\rms}{\textit{rms}}

\title[The 2016 outburst of GRS 1716--249]{The long outburst of the black hole transient GRS 1716--249 observed in the X-ray and radio band}

\author[T. Bassi et al.]{
T. Bassi$^{1,2,3}$,\thanks{E-mail: bassi@ifc.inaf.it}
M. Del Santo $^{1}$, 
A. D'A\`i$^{1}$,
S.E. Motta$^{4}$,
J. Malzac$^{3}$,
\newauthor
A. Segreto$^{1}$,
J.C.A. Miller-Jones$^{5}$,
P. Atri$^{5}$,
R.M. Plotkin$^{5}$,
\newauthor
T.M. Belloni$^{6}$,
T. Mineo$^{1}$,
A.K. Tzioumis$^{7}$
\\
$^{1}$ INAF -- Istituto di Astrofisica Spaziale e Fisica Cosmica di Palermo, Via Ugo La Malfa 153, I-90146 Palermo, Italia\\
$^{2}$ Universit\'a degli Studi di Palermo, Dipartimento di Fisica e Chimica, via Archirafi 36,  I-90123 Palermo, Italia\\
$^{3}$ IRAP, Universit\'{e} de Toulouse, CNRS, UPS, CNES, Toulouse, France\\
$^{4}$ Department of Physics, Astrophysics, University of Oxford, Denys Wilkinson Building, Keble Road,\\ 
 OX1 3RH Oxford, UK 0000-0002-6154-5843\\
$^{5}$ International Centre for Radio Astronomy Research -- Curtin University, GPO Box U1987, Perth, WA 6845, Australia\\
$^{6}$ INAF -- Osservatorio  Astronomico  di  Brera,  via  E.  Bianchi  46, I-23807,  Merate\\
$^{7}$ CSIRO Astronomy and Space Science, Australia Telescope National Facility, PO Box 76, Epping, NSW 1710, Australia
}

\date{Accepted ... Received ...; in original form ...}

\pubyear{2015}
\hypersetup{draft}

\begin{document}
\label{firstpage}
\pagerange{\pageref{firstpage}--\pageref{lastpage}}
\maketitle

\begin{abstract}
We present the spectral and timing analysis of X-ray observations performed on the Galactic black hole transient \grs~ during the 2016-2017 outburst. The source was almost continuously observed with the {\it Neil Gehrels Swift Observatory} from December 2016 until October 2017. The X-ray hardness ratio and timing evolution indicate that the source approached the soft state three times during the outburst, even though it never reached the canonical soft spectral state.
Thus, \grs~ increases the number of black hole transients showing outbursts with ``failed" state transition. During the softening events, XRT and BAT broadband spectral modeling, performed with thermal Comptonization plus a multicolor disc black-body, showed a photon index ($\Gamma$ < 2) and an inner disc temperature (\kti~ = 0.2-0.5\,keV) characteristic of the hard intermediate state. This is in agreement with the root mean square amplitude of the flux variability (\rms\ > 10$\%$). We find that, coherently with a scenario in which the disc moves closer to the compact object, the accretion disc inner radius decreases with the increase of the inner disc temperature, until a certain point when the temperature starts to increase at constant radius. This, in addition with the spectral analysis results, suggests that either the accretion disc reached the innermost stable circular orbit during the hard intermediate state or the hot accretion flow might re-condensate in an inner mini-disc. 
We report on the radio observations performed during the outburst finding that \grs~ is located on the radio-quiet ``outlier" branch of the radio/X-ray luminosity plane.
\end{abstract}

\begin{keywords}
X-rays: general -- accretion, accretion disc -- black hole -- X-rays: binaries -- stars: jet 
\end{keywords}



\section{Introduction}

Black hole transients (BHTs) are binary systems with a black hole (BH) accreting matter from a less evolved star. They alternate quiescent periods characterized by X-ray luminosities of L$_{\rm X} \sim$10$^{30-33}$\,\ergs~ and episodic outbursts during which the source can reach L$_{\rm X} \sim$10$^{36-39}$\,\ergs. 
During outbursts they can show different X-ray spectral states, characterized by different luminosities (low or high), spectral shapes (hard or soft; \citealt{zdz04,remillard06}) and timing properties \citep{belloni11, belloni16}.
The spectral evolution of BHTs can be described in terms of the Hardness-Intensity Diagram \citep[HID,][]{homan05} based on X-ray measurements, where the spectral states are positioned on different regions of a q-shaped pattern.\\
There are two main spectral states: the hard state (HS) and the soft state (SS), in which X-ray spectra are dominated by the hard X-ray emission and soft X-ray emission, respectively, and are usually explained in terms of changes in the geometry of the accretion flow onto the central object \citep{zdz00,done07}.
The HS corresponds to the right-hand branch of the HID, while the SS is located on the left-hand side. 

BHTs start their outbursts in the HS with a spectrum well described by a power law (typical photon index $\Gamma<2$) with a high-energy cut-off at $\sim$50-100\,keV. This is physically interpreted as thermal Comptonization due to Compton up-scattering of soft disc photons by a corona of hot electrons ($\sim$100\,keV) located close to the BH \citep{zdz04}. 
A weak soft X-ray thermal component with a temperature of $\sim$0.1-0.2\,keV is usually observed and interpreted as emission from an optically thick,  geometrically thin, accretion disc \citep{shakura73}. In this state, the accretion disc is thought to be truncated at large radii (roughly 100\,R$_{g}$) from the BH \citep{done07}.
\\
During the SS, BHTs spectra show a strong soft thermal component with inner disc temperature \kti $\sim$1\,keV. This emission is likely associated to the Shakura-Sunyaev disc extending down to the innermost stable circular orbit (ISCO). In addition, a steep power-law tail ($\Gamma$ > 2.5), which often extends to the high energies ($\geq$500\,keV), is observed. 
Usually, this component is interpreted as due to non-thermal Comptonisation, although its origin is still debated \citep{poutanen98, laurent99,niedzwiecki06}.\\
Two further spectral states, namely the Hard and Soft Intermediate States (HIMS and SIMS, respectively), are also defined (see, e.g., \citealt{belloni16}).  They are located along the horizontal branches in the HID, with spectral parameters in between those of the main states.\\
Not all BHTs reproduce the standard q-track pattern during their outbursts: an increasing number of sources show the transition HS-to-HIMS, but never make the transitions to the SIMS and SS \citep{capitanio09,ferrigno12,soleri13,delsanto16}. These outbursts, which do not evolve through the full set of spectral states, are called ``failed" state transition. The reasons why these sources fail to make transition to the soft state is still debated. 

The X-ray Power Density Spectra (PDS) of BHTs show different properties depending on the spectral state.  
The variability is typically quantified in terms of fractional root mean squared (\rms) variability. In the HS the fractional \rms\ can be as high as 30\% (see, e.g., \citealt{belloni16}), while the SS is characterized by very low variability levels (\rms\ < 5\%).
During the HIMS, the fractional \rms\ is observed to decrease  from $\sim$30\% to 10\% and correlates with the hardness \citep{munoz11}.
The most dramatic changes in the power density spectra occur at the transition between the HIMS and the SIMS \citep{belloni16}.  
This timing transition is extremely fast, as opposed to the smooth spectral changes usually observed across the HIMS to SIMS transition \citep{delsanto09, motta09}.
The fractional \rms\ in the SIMS is estimated to be between 7 and 10 per cent \citep{munoz11}.

In addition to the X/$\gamma$-ray phenomenology, the BHTs spectral states are also characterized by the production of different outflow  \citep{mirabel94, fender06}.
Jets, whose emission is observed mainly in the radio band, are coupled to the accretion flow, even though the nature of this connection is still unclear. 
The hard states of BHTs are typically associated with a \textit{compact} jet, characterized  by a flat or slightly-inverted radio spectrum (e.g \citealt{fender01,corbel02}). This is interpreted as self-absorbed synchrotron emission from steady jets, as also observed in low-luminosity AGNs \citep{blandford79}.  However, during the SS these compact jets are thought to be quenched \citep{fender99, corbel00}.
During the HS, a radio/X-ray connection has been observed in several sources in the form of a non-linear flux correlation F$_{\rm R} \propto$ F$_{\rm X}^{\rm a}$, where a $\sim$0.5-0.7, and F$_{\rm R}$ and F$_{\rm X}$ are the radio and X-ray fluxes, respectively (see e.g. \citealt{corbel03,gallo03}).
In recent years, a number of Galactic BHTs were found to have a steeper correlation index, i.e. $\sim$1.4, in the radio/X-ray luminosity plane \citep{coriat11,corbel13}.
They were called ``outliers", even though the increasing number of these sources  could in fact turn out to be the norm \citep{coriat11,motta18}. 
We refer to the ``outliers" also as radio-quiet since they can be radio fainter by 1-2 magnitudes around L$_{\rm X}\sim$10$^{36}$-10$^{37}$\,\ergs.   

\subsection{\grs}
The BH X-ray transient \grs~ (also known as \gro, Nova Oph 1993) was discovered in 1993 September with the {\it CGRO}/BATSE and {\it Granat}/SIGMA telescopes \citep{harmon93,ballet93}. 
At that time, it reached a maximum flux of 1.4\,Crab in the 20--100\,keV energy band.  
The optical counterpart was identified with the spectral type K (or possibly later) star V2293 Oph and a distance of $2.4 \pm 0.4$\,kpc was derived \citep{dellavalle94}. \cite{masetti96} estimated a lower limit for the compact object mass of 4.9\,M$_{\odot}$ (confirming the BH nature), an evolved companion star mass of 1.6\,M$_{\odot}$ and an orbital period of 14.7\,hr. The hydrogen column density along the line of sight was estimated being $\sim$4 $\times$ 10$^{21}$\,atoms\,cm$^{-2}$ \citep{tanaka93}.
In 1993 October, the source was also observed in radio by the Very Large Array (VLA), showing a flat spectrum \citep{dellaValle93,dellavalle94}. A peculiar outburst showing five sawtooth-like shapes in the X-ray light curve was observed in 1995 \citep{hjellming96}. Simultaneous radio observations detected a synchrotron radio flare, implying the presence of relativistic particles and magnetic fields.\\
\grs~ was detected again in outburst on 2016 December 18  by {\it MAXI} after more than twenty years in quiescence \citep{negoro16,masumitsu16}. Soon after, a {\it Neil Gehrels Swift Observatory} \citep[hereafter \swift,][]{gehrels04} XRT monitoring campaign \citep{delsanAtel} was triggered and a preliminary spectral study showed a hard X-ray spectrum. On 2017 February 10,  an \integral~ ToO of 90\,ks showed that the source was in hard state \citep{delsanAtel2}. \integral~ results combined with simultaneous radio and infrared observations will be presented in a further paper. 

In this paper, we present \swift~ spectral and timing analysis results and radio observations performed during the whole 2016-2017 outburst of \grs~ which lasted about ten months.

\section{X-ray Observations and Data Reduction}
\grs~ was observed both with the X-Ray Telescope \citep[XRT,][]{burrows05} and the Burst Alert Telescope \citep[BAT,][]{barthelmy05} on board \swift. XRT works in the soft X-ray band (0.2--10\,keV) in four different operating modes depending on the source count rate \citep[for more details see ][]{burrows05}. 
The Windowed Timing (WT) mode is used to obtain high time resolution light curves (1.7\,ms) and it is used to observe very bright sources (1--600\,mCrab).
BAT is a highly sensitive coded mask instrument. It can daily observe up to 80\% of the whole sky in survey mode in the 15--150\,keV band, aiming to catch gamma-ray burst events.\\
We retrieved and analyzed all XRT and BAT observations performed in the period 2016 December-2017 October.
We used XSPEC v. 12.9.1 for X-ray spectral analysis. 
Errors on spectral parameters are given at the 90$\%$ confidence level.

\subsection{XRT} \label{xrt}
The XRT monitoring campaign of \grs~ was performed in WT 
observing mode from 2017 January 28 (MJD 57781) to 2017 October 20 (MJD 58046), with target IDs 34924 and 88233 (the observation log is provided in Table \ref{log}). In observations $\#$046, $\#$059 and $\#$067 the source was out of the part of the CCD used in WT mode and we therefore excluded these data from the analysis.
We extracted the 0.2--10\,keV count rate from each  observation (see Tab. \ref{log}) corrected for instrumental artifacts (i.e. bad columns on the CCD) using the on-line products generator\footnote{\url{http://www.swift.ac.uk/user\_objects/}} \citep{evans07, evans09}.
With the same tool, we extracted pointing by pointing the count rate in 0.5--2\,keV and 2--10\,keV energy ranges in order to study the Hardness-Ratio (HR), defined as the ratio of the hard count rate over the soft one (see Sec. \ref{results} and Fig. \ref{lc_hr}).
Then, we reprocessed the XRT data using the FTOOLS software package (HEASoft v.6.20) and the \swift\ Calibration Database (CALDB; release 20160609). We ran the task {\it xrtpipeline}, selecting only events with grade 0. This filter is applied in the analysis of bright sources in order to reduce the spectral distortion at low energies ($\lesssim$ 1.0\,keV) observed in WT mode\footnote{\url{http://www.swift.ac.uk/analysis/xrt/digest\_cal.php}}. Then we selected a circle of 20 pixel (1 pixel = 2.357$''$) radius centred on the source position to extract the spectra using the task {\it xrtproducts}. When the source was on the edge of the CCD, we selected a region with the largest size possible; in pointings $\#$054, $\#$058 and $\#$060, radii of 5, 8 and 4 pixels respectively were chosen. The background was extracted from a region of the same size of the source region located away from the source. 

According to \cite{romano06} the pile-up correction should be applied for sources with count rate above 100\,\cnt. However, we have evaluated that pointings with count rate between 90--100\,\cnt~ were still affected by pile-up\footnote{\url{http://www.swift.ac.uk/analysis/xrt/pileup.php}}.
Thus, the pile-up correction was carried out for observations with count rate (values plus uncertainties) $\geq$ 90\,\cnt~ (see Tab. \ref{log}).
For these observations we extracted the spectra from an annular region centred on the source with an outer radius of 20 pixels and an inner radius of 3 pixels. 
In order to apply the $\chi^2$ statistics we grouped the channels for having at least 50 counts per energy bin.
We computed power spectra in the energy band 0.4--10.0\,keV from each XRT observation, using custom software written in IDL\footnote{GHATS, \url{http://www.brera.inaf.it/utenti/belloni/GHATS\_Package/Home.html}}. We used $\approx$29-s long intervals and a Nyquist frequency of $\approx$280\,Hz. We averaged the PDS in order to obtain one PDS per observation, which we normalized according to \cite{Leahy1983}. We measured the fractional \rms~in the frequency range 0.035--10\,Hz. After MJD 58000 (2017 September 4) the source count rate was too low ($\lesssim$ 5 counts/s) to allow us to measure the fractional \rms, the upper limit to which is essentially 100\%. For this reason, we do not report any fractional \rms~value after this date. 

\subsection{BAT}
\grs\ was almost daily observed in survey mode with BAT from 2016 December 1 (MJD 57723). We processed the BAT data, available from the HEASARC public archive, using the \texttt{BAT-IMAGER} software \citep{segreto10}.
This software performs screening, mosaicking and source detection, and produces scientific products for any detected source. We used the official BAT spectral redistribution matrix. \\
We extracted the light curves in two energy bands (15--30\,keV and 30--90\,keV) in order to compute the HR, defined as the count rate ratio [30--90]\,keV/[15--30]\,keV, with one-day time interval. \\
Spectra were extracted in 30 channels with logarithmic binning in the energy range 15--185\,keV. Due to the poor statistics of the hard X-ray emission at the beginning of the outburst, data from MJD 57723 until MJD 57739 were excluded.\\

\onecolumn
\begin{center}
{\renewcommand\arraystretch{1.2}
\begin{longtable}{lclccc}
\caption{XRT pointings of \grs~ analysed in this work. The columns are: (1) sequence number, (2) start time in Terrestrial Time (TT), (3) exposure time, (4) 0.2--10 keV count rate and errors obtained using the on-line products generator \citep{evans07, evans09}. In columns (5) and (6) we report the photon index and the reduced chi squared, respectively, obtained by fitting spectra with an absorbed power-law model (see Sec. \ref{results}).\\
$^{\dag}$ Observations not used since \grs~ was out of the part of the CCD used in WT mode.\\
$^{\ast}$ Pile-up correction has been applied.\\
$^{\diamond}$ Spectrum has been extracted in a region of radius smaller than 20 pixels since the source was on the edge of the CCD.\\
$^{\ddag}$ Spectral fits need an additional component (i.e. {\sc diskbb}, see Sec. \ref{broadband}) to the simple power-law.
 \\
}  
\label{log} \\
\hline 

\textbf{Seq} & \textbf{Start Time} & \textbf{Exposure} & \textbf{Count Rate} &\textbf{$\Gamma$ } &\textbf{\chir(dof)} \tabularnewline 
\textbf{$\#$} &  &\textbf{(s)} & \textbf{(\cnt)} &  & \tabularnewline
(1) & (2) & (3) & (4) & (5) & (6) \tabularnewline
\hline
\tabularnewline
\endfirsthead

\multicolumn{3}{c}%
{{\bfseries \tablename\ \thetable{} -- continued from previous page}} \\
\hline \textbf{Seq} & \textbf{Start Time} & \textbf{Exposure} & \textbf{Average Rate} &\textbf{$\Gamma$ } &\textbf{\chir(dof)} \tabularnewline 
\textbf{$\#$} &  &\textbf{(s)} & \textbf{(\cnt, 0.2--10\,keV)} &  & \tabularnewline
(1) & (2) & (3) & (4) & (5) & (6) \tabularnewline

\hline 
\tabularnewline
\endhead
\\
\hline 
\\
\multicolumn{6}{c}{{Continued on next page}} \\ 
\\
\hline
\endfoot

\hline \hline
\endlastfoot

001$^{\ast}$	 &	2017-01-28 00:08:01 &	992  &	 93.1$\pm$0.4	&  1.55$^{+0.03}_{-0.03}$  &  1.08(342)  	\tabularnewline
002$^{\ast}$	 &	2017-01-29 19:08:57 &	999  &	 102$\pm$3  	&  1.57$^{+0.04}_{-0.05}$  &  0.93(315)  	\tabularnewline
003$^{\ast}$	 &	2017-01-31 00:05:15 &	683  &	 95$\pm$2  	&  1.58$^{+0.04}_{-0.04}$  &  1.10(248) 	\tabularnewline
004$^{\ast}$	 &	2017-02-02 17:31:57 &	990  &	 100$\pm$4  	&  1.60$^{+0.03}_{-0.03}$  &  1.04(372)  	\tabularnewline
005$^{\ast}$	 &	2017-02-04 18:56:10 &	924  &	 100.2$\pm$0.3	&  1.58$^{+0.03}_{-0.03}$  &  1.03(373)  	\tabularnewline
006$^{\ast}$	 &	2017-02-06 10:59:09 &	305  &	 96$\pm$1  	&  1.53$^{+0.06}_{-0.05}$  &  0.95(138)  	\tabularnewline
007$^{\ast}$	 &	2017-02-08 04:15:42 &	946  &	 97.9$\pm$0.3	&  1.59$^{+0.03}_{-0.03}$  &  1.12(384)  	\tabularnewline
008$^{\ast}$	 &	2017-02-09 02:19:11 &	1478 &	 80$\pm$12 	&  1.62$^{+0.03}_{-0.03}$  &  1.07(432)   	\tabularnewline
009$^{\ast}$	 &	2017-02-09 05:43:11 &	1282 &	 87$\pm$11 	&  1.57$^{+0.03}_{-0.03}$  &  1.02(365) 	\tabularnewline
010$^{\ast}$	 &	2017-02-09 09:01:44 &	1451 &	 97.4$\pm$0.3	&  1.62$^{+0.03}_{-0.02}$  &  1.03(452)  	\tabularnewline
011$^{\ast}$	 &	2017-02-09 13:27:07 &	1287 &	 102$\pm$1  	&  1.64$^{+0.03}_{-0.03}$  &  1.14(400)  	\tabularnewline
012$^{\ast}$	 &	2017-02-09 18:11:07 &	1360 &	 103$\pm$3  	&  1.64$^{+0.03}_{-0.03}$  &  0.89(364)  	\tabularnewline
013$^{\ast}$	 &	2017-02-09 21:27:58 &	1367 &	 103$\pm$0.7	&  1.62$^{+0.03}_{-0.03}$  &  1.00(425)  	\tabularnewline
014$^{\ast}$	 &	2017-02-11 07:11:47 &	949  &	 101$\pm$3  	&  1.57$^{+0.04}_{-0.03}$  &  0.99(299) 	\tabularnewline
015$^{\ast}$	 &	2017-02-12 13:17:33 &	1209 &	 81$\pm$12 	&  1.57$^{+0.03}_{-0.03}$  &  1.06(431)  	\tabularnewline
016$^{\ast}$	 &	2017-02-14 08:42:43 &	921  &	 87$\pm$4  	&  1.55$^{+0.03}_{-0.03}$  &  1.10(322)  	\tabularnewline
017$^{\ast}$	 &	2017-02-14 19:35:14 &	975  &	 91.6$\pm$0.4	&  1.60$^{+0.03}_{-0.03}$  &  0.99(358)  	\tabularnewline
018$^{\ast}$	 &	2017-02-16 03:30:55 &	997  &	 90.2$\pm$0.3	&  1.59$^{+0.03}_{-0.03}$  &  1.17(357)  	\tabularnewline
019$^{\ast}$	 &	2017-02-22 13:56:26 &	1734 &	 92$\pm$5  	&  1.57$^{+0.03}_{-0.03}$  &  1.05(398) 	\tabularnewline
020$^{\ast}$	 &	2017-02-24 09:00:33 &	966  &	 100.3$\pm$0.3	&  1.56$^{+0.03}_{-0.03}$  &  1.19(362)  	\tabularnewline
021$^{\ast}$	 &	2017-02-26 15:27:03 &	934  &	 100.1$\pm$0.3	&  1.60$^{+0.03}_{-0.03}$  &  1.05(348)   	\tabularnewline
022$^{\ast}$	 &	2017-03-09 12:48:10 &	1063 &	 97.4$\pm$0.4	&  1.63$^{+0.03}_{-0.03}$  &  1.04(375)  	\tabularnewline
024$^{\ast}$	 &	2017-03-15 18:46:20 &	985  &	 90.2$\pm$0.3	&  1.59$^{+0.03}_{-0.03}$  &  1.08(348)  	\tabularnewline
025$^{\ast}$	 &	2017-03-21 20:16:46 &	959  &	 98.7$\pm$0.3	&  1.67$^{+0.03}_{-0.03}$  &  0.92(384)  	\tabularnewline
026$^{\ast}$	 &	2017-03-27 22:36:50 &	940  &	 124$\pm$11 	&  1.69$^{+0.03}_{-0.03}$  &  0.92(333)  	\tabularnewline
027$^{\ast}$	 &	2017-04-02 09:07:26 &	960  &	 128$\pm$6  	&  1.76$^{+0.03}_{-0.03}$  &  1.09(389)  	\tabularnewline
029$^{\ast\ddag}$	 &	2017-04-07 08:50:41 &	1664 &	 131$\pm$1  	&  1.85$^{+0.02}_{-0.02}$  &  1.21(511) 	\tabularnewline
030$^{\ast\ddag}$	 &	2017-04-08 23:11:36 &	965  &	 131.0$\pm$0.7	&  1.93$^{+0.05}_{-0.03}$  &  1.33(413)  	\tabularnewline
031$^{\ast\ddag}$	 &	2017-04-10 21:12:42 &	1933 &	 143$\pm$6  	&  2.01$^{+0.02}_{-0.02}$  &  1.19(514)   	\tabularnewline
032$^{\ast\ddag}$	 &	2017-04-11 03:50:28 &	727  &	 147$\pm$3  	&  2.06$^{+0.03}_{-0.03}$  &  1.20(383)  	\tabularnewline
033$^{\ast\ddag}$	 &	2017-04-12 11:26:48 &	1852 &	 137$\pm$3  	&  1.98$^{+0.02}_{-0.02}$  &  1.23(508)  	\tabularnewline
034$^{\ast\ddag}$	 &	2017-04-14 00:37:03 &	929  &	 123$\pm$7  	&  1.92$^{+0.03}_{-0.03}$  &  1.13(351)   	\tabularnewline
035$^{\ast}$	 &	2017-04-21 01:39:30 &	444  &	 98.1$\pm$0.5	&  1.75$^{+0.04}_{-0.04}$  &  1.07(236)  	\tabularnewline
036     	 &	2017-05-05 17:42:16 &	1152 &	 83$\pm$0.3	&  1.72$^{+0.02}_{-0.02}$  &  1.17(525)  	\tabularnewline
037     	 &	2017-05-11 01:28:31 &	1258 &	 79.2$\pm$0.3	&  1.71$^{+0.02}_{-0.02}$  &  1.03(538)  	\tabularnewline
038$^{\ast\ddag}$	 &	2017-05-22 22:52:31 &	839  &	 100.3$\pm$0.4	&  1.90$^{+0.03}_{-0.03}$  &  1.11(333) 	\tabularnewline
039$^{\ast}$	 &	2017-05-26 17:55:22 &	724  &	 91.8$\pm$0.5	&  1.79$^{+0.04}_{-0.04}$  &  1.15(261) 	\tabularnewline
040     	 &	2017-06-02 12:29:01 &	354  &	 78.6$\pm$0.5	&  1.79$^{+0.03}_{-0.03}$  &  0.99(295) 	\tabularnewline
041     	 &	2017-06-07 00:25:55 &	1059 &	 74.8$\pm$0.3	&  1.75$^{+0.02}_{-0.02}$  &  1.20(486)  	\tabularnewline
042     	 &	2017-06-12 04:46:37 &	3354 &	 75.4$\pm$0.9	&  1.81$^{+0.01}_{-0.01}$  &  1.15(641) 	\tabularnewline
043     	 &	2017-06-17 02:58:56 &	1058 &	 76$\pm$1  	&  1.82$^{+0.02}_{-0.02}$  &  1.17(475) 	\tabularnewline
044     	 &	2017-06-23 21:29:38 &	1033 &	 74.0$\pm$0.3	&  1.91$^{+0.02}_{-0.02}$  &  1.13(464)  	\tabularnewline
045$^{\ast\ddag}$	 &	2017-06-30 20:57:38 &	437  &	 91$\pm$4  	&  2.13$^{+0.04}_{-0.04}$  &  1.29(267)  	\tabularnewline
046$^{\dag}$   	 &	2017-07-09 18:39:24 &	964  &	 -              &           	  -        &  -                 \tabularnewline   
047$^{\ddag}$     	 &	2017-07-16 03:43:26 &	978  &	 66.9$\pm$0.3	&  2.12$^{+0.02}_{-0.02}$  &  1.31(413)  	\tabularnewline
048$^{\ast\ddag}$	 &	2017-07-23 11:12:15 &	988  &	 100.1$\pm$0.3	&  2.64$^{+0.04}_{-0.04}$  &  1.35(295) 	\tabularnewline
049$^{\ast\ddag}$	 &	2017-07-25 15:33:47 &	980  &	 106.0$\pm$0.3	&  2.68$^{+0.04}_{-0.04}$  &  1.30(286) 	\tabularnewline
050$^{\ast\ddag}$	 &	2017-07-26 17:10:39 &	1491 &	 110.6$\pm$0.3	&  2.79$^{+0.03}_{-0.03}$  &  1.34(349) 	\tabularnewline
051$^{\ddag}$     	 &	2017-07-29 20:00:17 &	972  &	 80.6$\pm$0.3	&  2.57$^{+0.02}_{-0.02}$  &  1.41(404) 	\tabularnewline
052$^{\ddag}$     	 &	2017-07-30 18:30:24 &	973  &	 78.3$\pm$0.3	&  2.55$^{+0.02}_{-0.02}$  &  1.42(397) 	\tabularnewline
053$^{\ddag}$     	 &	2017-07-31 22:58:09 &	919  &	 74.7$\pm$0.3	&  2.53$^{+0.03}_{-0.03}$  &  1.46(388)	        \tabularnewline
054$^{\diamond\ddag}$ &	2017-08-06 19:19:08 &	863  &	 20.8$\pm$0.2   &  1.95$^{+0.04}_{-0.04}$  &  0.99(229)      	\tabularnewline
055$^{\ddag}$     	 &	2017-08-11 06:27:11 &	817  &	 28.7$\pm$0.2	&  1.87$^{+0.04}_{-0.04}$  &  0.99(263) 	\tabularnewline
056     	 &	2017-08-13 09:08:40 &	1081 &	 23.1$\pm$0.2	&  1.81$^{+0.04}_{-0.03}$  &  1.17(277) 	\tabularnewline
057     	 &	2017-08-15 05:46:59 &	1058 &	 21.4$\pm$0.1	&  1.74$^{+0.04}_{-0.04}$  &  0.95(252)  	\tabularnewline
058$^{\diamond}$ &	2017-09-08 03:38:00 &	1667 &	 4.24$\pm$0.07  &  1.58$^{+0.07}_{-0.06}$  &  0.93(107)         \tabularnewline
059$^{\dag}$   	 &	2017-09-11 01:47:00 &	1529 &	 -  	 	&                 -        &  -                 \tabularnewline
060$^{\diamond}$ &	2017-09-15 00:05:15 &	1226 &	 2.85$\pm$0.08  &  1.52$^{+0.11}_{-0.10}$  &  0.89(51)          \tabularnewline
061     	 &	2017-09-17 00:05:15 &	1106 &	 3.31$\pm$0.07  &  1.58$^{+0.11}_{-0.10}$  &  0.95(45)          \tabularnewline
062     	 &	2017-09-21 01:07:13 &	1639 &	 3.24$\pm$0.06  &  1.55$^{+0.07}_{-0.07}$  &  0.95(94)          \tabularnewline
063     	 &	2017-10-01 00:10:50 &	1425 &	 4.09$\pm$0.06  &  1.47$^{+0.08}_{-0.07}$  &  0.99(84)          \tabularnewline
064     	 &	2017-10-05 01:25:00 &	1591 &	 4.29$\pm$0.07  &  1.56$^{+0.09}_{-0.08}$  &  1.29(69)          \tabularnewline 
065     	 &	2017-10-10 00:57:59 &	1372 &	 3.61$\pm$0.06  &  1.63$^{+0.10}_{-0.09}$  &  1.05(68)          \tabularnewline 
066     	 &	2017-10-15 00:33:11 &	1477 &	 3.66$\pm$0.06  &  1.61$^{+0.08}_{-0.08}$  &  1.27(75)          \tabularnewline 
067$^{\dag}$   	 &	2017-10-20 11:21:50 &	1388 &	 -              &         	  -        &  -                 \tabularnewline 
\tabularnewline                                                                                                                                                                                 
  \hline                                                                                                                
\tabularnewline                                                                                                         
ID 88233001$^{\ast\ddag}$       &	2017-07-28 16:51:27 &	3716 &	 100.0$\pm$0.2	&  2.75$^{+0.02}_{-0.02}$  &  1.57(481)         \tabularnewline                                               
\tabularnewline
\end{longtable}}
\end{center}
\twocolumn 

\section{Radio observations and data reduction}
\label{sec:radio}
\subsection{ATCA}

We observed \grs\ with the Australia Telescope Compact Array (ATCA) from 18:52--21:00 UTC on 2017 February 9, under program code C2538.  We observed simultaneously in two 2-GHz bands, centred at 5.5 and 9.0\,GHz, with the array in its extended 6D configuration, with a maximum baseline of 6\,km.  We used PKS 1934$-$638 as our amplitude and bandpass calibrator, and PKS 1710$-$269 (2.4$^{\circ}$ from \grs) as our complex gain calibrator, adopting a cycle time of 16.5\,min, with 15\,min spent on \grs\ and 1.5\,min on the calibrator in each cycle.  Data reduction was performed according to standard procedures in the Multichannel Image Reconstruction, Image Analysis, and Display \textsc{Miriad} software package \citep{Sault95}.  The calibrated data were imported into the Astronomical Image Processing System \citep[{\sc aips};][]{greisen03} for imaging, where we used pure uniform weighting to minimise the elongation of the synthesised beam that arose from the sparse {\it uv}-coverage.  Following deconvolution, the source flux density was measured by fitting a point source in the image plane.

\subsection{LBA}

We observed \grs\ with the Australian Long Baseline Array (LBA) on three occasions; 2017 February 21, April 22 and August 13, under program code V447.  The first two observations were taken outside formal LBA sessions, with a minimal array of just four antennas.  However, the August 13 observation was taken during an official LBA run, allowing us to include ten stations in the array.  Details of the different observations are given in Table~\ref{tab:lbalog}.

Data were correlated using the DiFX software correlator \citep{deller11}, and reduced using standard procedures within {\sc aips}.  Given the faintness of the target, we observed in phase referencing mode, nodding between \grs\ and the nearby phase reference calibrator J1711-2509, located $1.87^{\circ}$ from \grs.  In each cycle, we spent 90\,s on the calibrator and 210\,s on the target.  Every fifth target scan was substituted for a scan on a nearby check source, J1713-2658.  Amplitude calibration was performed using the system temperatures recorded at the individual stations where available, and nominal values in other cases.  We used a bright fringe finder source to perform instrumental phase and delay corrections, and subsequently to calibrate the instrumental frequency response.  We hybrid mapped the phase reference source, before using it to derive the time-dependent amplitude, phase, delay and rate solutions, which were applied to \grs.  After imaging the target source, its flux density was determined by fitting a point source in the image plane, since in no case was it observed to be extended.  The loss of one antenna (Cd) for the majority of the first observation (2017 February 21) meant that imaging with three antennas could not reliably reproduce the source structure, so for this epoch we fit the data with a point source in the {\it uv}-plane.

\begin{table*}
	\centering
	\caption{LBA observations of \grs.  Station codes are At (the phased-up ATCA); Cd (Ceduna); Ho (Hobart); Ke (Katherine); Mp (Mopra); Pa (Parkes); Td (the 34m DSS36 antenna at Tidbinbilla);  Ti (the 70m DSS43 antenna at Tidbinbilla); Ww (the Warkworth 12m antenna); Yg (Yarragadee).}
	\label{tab:lbalog}
	\begin{tabular}{lcccl} 
		\hline
		Program code & Date, Time (UT) & MJD & Bandwidth & Array\\
		\hline
		V447D & 2017-02-21, 17:18--02:30 & $57805.91\pm0.19$ & $1\times64$\,MHz & At-Cd-Ho-Mp\\
		V447E & 2017-04-22, 11:56--23:00 & $57865.73\pm0.23$ & $4\times16$\,MHz & At-Cd-Ho-Mp\\
		V447F & 2017-08-13, 06:56--16:00 & $57978.48\pm0.19$ & $4\times16$\,MHz & At-Cd-Ho-Ke-Mp-Pa-Td-Ti-Wa-Yg\\
		\hline
	\end{tabular}
\end{table*}

\subsection{VLA}
\label{sec:vla}

We observed \grs\ with the Karl G.\ Jansky Very Large Array (VLA) from 05:48--06:35 UTC on 2017 August 12, under program code SI0462.  The array was in its moderately compact C configuration, and we split the antennas into two subarrays of 14 and 12 antennas, each of which observed simultaneously in two 1-GHz basebands, centred at 5.25 and 7.45\,GHz, and 8.8 and 11.0\,GHz, respectively.  \grs\ was setting during the observations, with an elevation decreasing from 19$^{\circ}$ to 14$^{\circ}$.  The weather was poor, with overcast skies, light rain and thunderstorms over the array leading to poor atmospheric phase stability.  Combined with the low elevation, this led to significant phase decorrelation, even at frequencies as low as 5\,GHz.

We used 3C\,286 as our flux and bandpass calibrator, and J1751$-$2524 as our complex gain calibrator.  We conducted two 17.5-minute scans on \grs\, each of which was bracketed by a 1-minute scan on the complex gain calibrator.  Data reduction was carried out following standard procedures within the Common Astronomy Software Application \citep[{\sc casa};][]{mcmullin07}.  We imaged the calibrated data using Briggs weighting, with a robust parameter of 1.  The image noise levels we achieved were significantly higher than expected theoretically (by factors of 2--4, increasing with frequency).  We attribute this to the higher system temperatures arising from the poor weather.  \grs\ was significantly detected at all frequencies, and we determined the source flux density by fitting it with a point source in the image plane.

Given the weather conditions at the array, we used the middle scan on the complex gain calibrator to quantify the extent of the phase decorrelation.  We re-reduced the data, using only the first and last calibrator scans to solve for the complex gain solutions, treating the middle calibrator scan as a second target.  A comparison of the calibrator flux densities when treated as a target (interpolating the gain solutions from the initial and final scans) showed that the peak amplitudes were lower by 9.6, 14.3, 25.7 and 32.3\% at 5.25, 7.45, 8.8 and 11.0\,GHz, respectively.  We therefore corrected the fitted flux densities of \grs\ by these factors to derive the final flux densities, as detailed in Table~\ref{tab:radio}.

\begin{table*}
	\centering
	\caption{Measured radio flux densities of \grs.  Quoted uncertainties are statistical only.  Nominal systematic uncertainties are of order 5\% for the VLA and ATCA, and at least 10\% for the LBA.  VLA flux densities have been corrected for the measured phase decorrelation, as described in Section~\ref{sec:vla}.}
	\label{tab:radio}
	\begin{tabular}{lccclc} 
		\hline
		Array & Date & MJD & Frequency & Flux density & Spectral Index\\
        & & & (GHz) & (mJy)&\\
		\hline
        ATCA & 2017 Feb \phantom{1}9 & $57793.835\pm0.040$ & 5.5 & $3.28\pm0.05$ & $-0.15\pm0.19$\\
        ATCA & 2017 Feb \phantom{1}9 & $57793.835\pm0.040$ & 9.0 & $3.04\pm0.03$\\
        \hline
        LBA & 2017 Feb 21 & $57805.911\pm0.192$ & 8.4 & $1.28\pm0.15$ & -\\
        \hline
        LBA & 2017 Apr 22 & $57865.729\pm0.225$ & 8.4 & $1.13\pm0.11$ & -\\
        \hline
        VLA & 2017 Aug 12 & $57977.256\pm0.014$ & 5.25 & $0.63\pm0.04$ & $-0.07\pm0.19$\\
        VLA & 2017 Aug 12 & $57977.256\pm0.014$ & 7.45 & $0.48\pm0.05$ &\\
        VLA & 2017 Aug 12 & $57977.256\pm0.014$ & 8.8 & $0.68\pm0.11$ &\\
        VLA & 2017 Aug 12 & $57977.256\pm0.014$ & 11.0 & $0.70\pm0.14$ &\\
        \hline
        LBA & 2017 Aug 13 & $57978.479\pm0.188$ & 8.4 & $0.29\pm0.04$ & -\\
		\hline
	\end{tabular}
\end{table*}

\begin{figure*}
   \centering
   	\includegraphics[width=15cm]{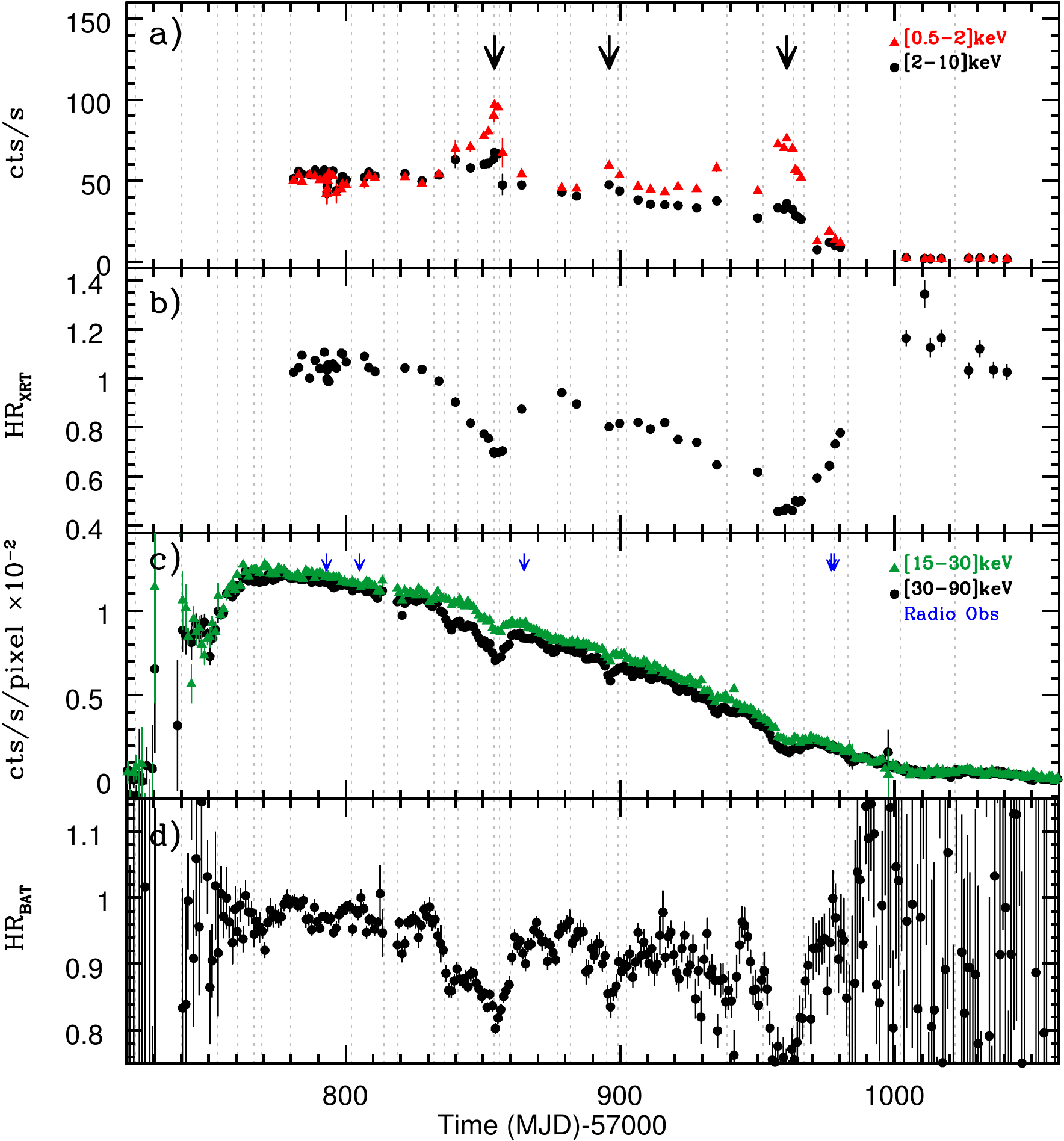}
   \caption{XRT light curves in the 0.5--2\,keV and 2--10\,keV energy ranges, extracted by pointing (panel {\it a}), plotted with the corresponding hardness ratio (panel {\it b}).
   In panel {\it c}, we show the 15--30\,keV and 30--90\,keV light curves observed by BAT, with a 1-day binning time, and the related hardness ratio (panel {\it d}). Dashed lines indicate the GTI considered for the broadband spectral analysis. In the XRT light curve,  three strong peaks at MJD 57857.03, 57906.52 and 57964.77 (marked with black arrows) in correspondence to dips in the BAT light curve and in the hardness ratios have been observed.
This indicates softening in the X-ray spectra.
   }
  \label{lc_hr}
\end{figure*}

\section{Results} 
\label{results}

The XRT and BAT count rate and HR evolution of \grs~ are shown in Figure \ref{lc_hr}. 
We noted three peaks in the XRT light curves (panel {\it a} in Fig. \ref{lc_hr}) which reach their maximum on MJD 57854.16 ($\#032$) with a total of 143\,\cnt, on MJD 57895.95 ($\#038$) with 97\,\cnt~ and on MJD 57960.72 ($\#050$) with 103\,\cnt. They correspond to dips in the BAT light  curves (panel {\it c} in Fig. \ref{lc_hr}) and in both the HRs (panels {\it b} and {\it d} in Fig. \ref{lc_hr}), and clearly indicate a spectral softening. The third peak corresponds to the softest observed episode: the 2--10\,keV count rate was lower than the 0.5--2\,keV count rate by a factor of two, whereas it was of 30$\%$ and 20$\%$ for the first two peaks.

In Figure \ref{hi_xrt} (top panel) we show the XRT Hardness-Intensity Diagram (HID) of \grs.
We plotted the total 0.5--10\,keV count rate versus  the count rate ratio 2--10\,keV/0.5--2\,keV. 
It is worth noting that due to the late XRT trigger with respect to the outburst beginning (the XRT monitoring started a few days after the hard X-ray peak), the HS right-hand branch in the HID was missed and the \grs\ q-track starts from the bright HS.
The outburst evolved towards the  intermediate states (IMS) on the horizontal branch. 
The three soft X-ray peaks in the XRT light curves are shown on the HID with a purple triangle, a green diamond and a magenta square.
The X-ray spectra of \grs~are observed to soften and harden twice along the IMS branch, until it reaches the softest observed state (magenta square), then the flux starts to decrease and the source simultaneously becomes harder along a diagonal track.
We noted that the shape of this HID is different from that observed in the majority of BHTs \citep[e.g. GX 339-4,][]{belloni06, dunn10}. Indeed there is not a clear SS branch on the left side of the diagram as observed in H 1743--322 and MAXI J1836--194 by \cite{capitanio09} and \cite{ferrigno12}, respectively.

\subsection{\textit{XRT} temporal variability}
\begin{figure*}
   \centering
   \includegraphics[width=15cm]{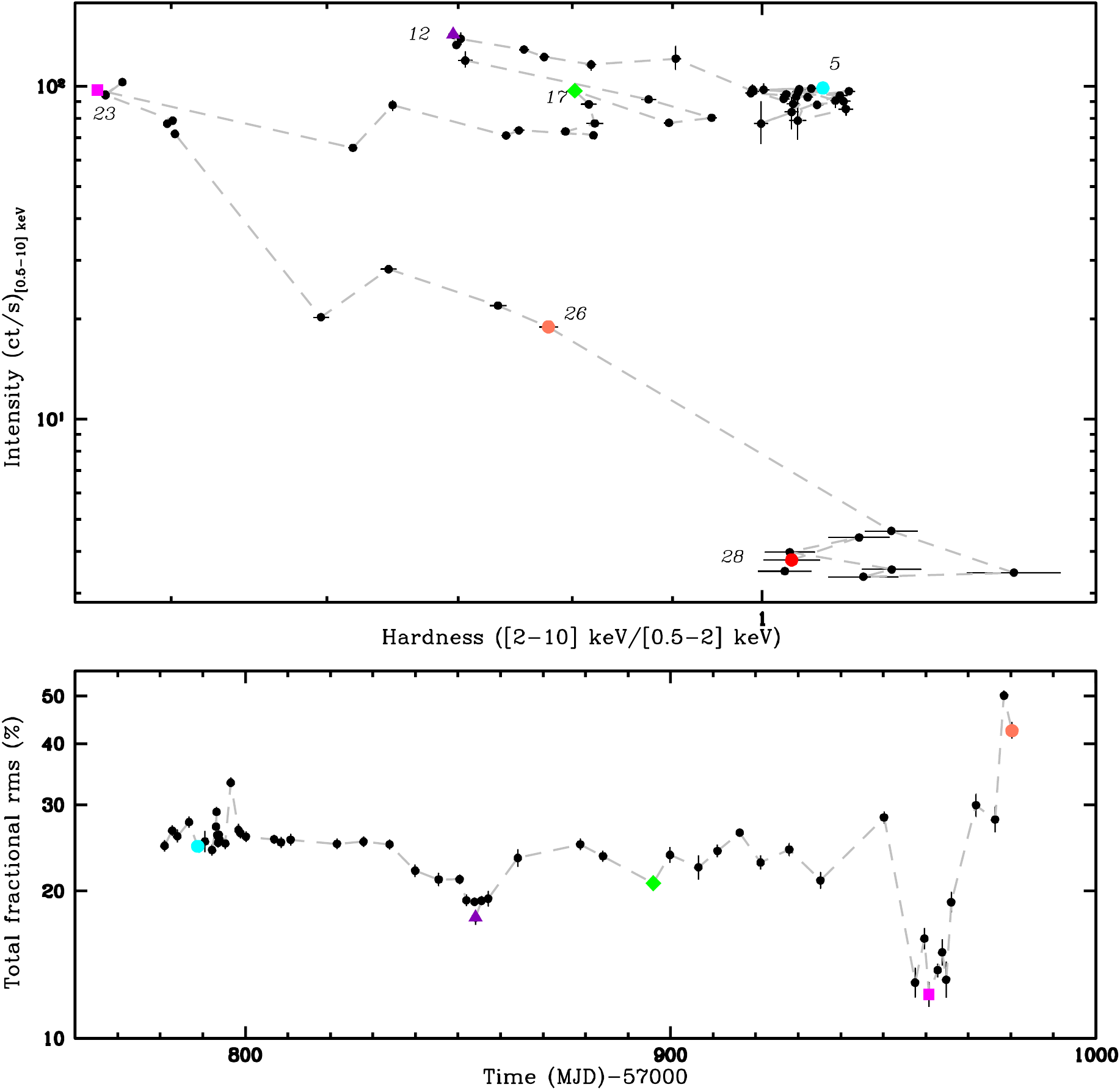}

   \caption{{\it Top panel:} Hardness-Intensity Diagram. The colored dots are associated with the broadband spectra shown in Fig. \ref{fit_spectra}. The outburst has been observed by XRT when \grs\ was at the top-right side of the pattern (cyan dot, broadband spectrum {\it 5}), then it evolves along the horizontal branch (purple triangle, spectrum {\it 12}). It reaches the softest state (magenta square, spectrum {\it 23}) and then takes a diagonal trajectory (orange dot, spectrum {\it 26}) on its return to the hard state (red dot, spectrum {\it 28}), before going in quiescence. The magenta square, purple triangle, and green diamond correspond to the softest points in each of the three softening episodes. {\it Lower panel:} XRT fractional \rms\ evolution. The soft points observed in the HID correspond to fractional \rms~ values typical of the HIMS \citep[10-30$\%$, ][]{munoz11}. }
              \label{hi_xrt}
\end{figure*}

In BHTs the fractional \rms\  of the variability is related to the spectral state \citep{munoz11}.
Therefore, we studied the evolution of the fractional \rms\ during the \grs\ outburst (Fig. \ref{hi_xrt} bottom panel). We noted that the fractional \rms~ measured from the XRT data is not directly comparable with those from \rxte/PCA \citep{munoz11}, because of the different energy bands which the two instruments are sensitive to.
However, the overall behaviour of the fractional \rms~ measured with XRT is expected to be similar to that would be measured by \rxte~ (with a difference in normalization).
In the first XRT observations \grs\ showed fractional \rms\ between 25 and 30$\%$ (e.g. cyan dot), typical of the bright HS at the beginning of the HIMS branch. 
Then, after MJD 57830, the fractional \rms\ started to decrease down to a value of 17$\%$ (MJD 57854), which corresponds to the purple triangle in the HID (Fig. \ref{hi_xrt} top panel) and to the first dip in the HRs (Fig. \ref{lc_hr}, b and d).
Afterwards the fractional \rms\ increased again up to values of about 25$\%$, then a second dip (green diamond) with fractional \rms\ of 20$\%$ at MJD 57896 (corresponding to the second softening) occurred.
Simultaneously with the third HR softening, the fractional \rms\ decreased down to 12$\%$ (magenta square, at MJD 57961).
Finally, it rose back to 40$\%$ in the final part of the outburst, indicating that the source returned to the HS.
 
\subsection{XRT spectral analysis} \label{xrt}

\begin{figure*}
   \centering
    \includegraphics[width=15cm]{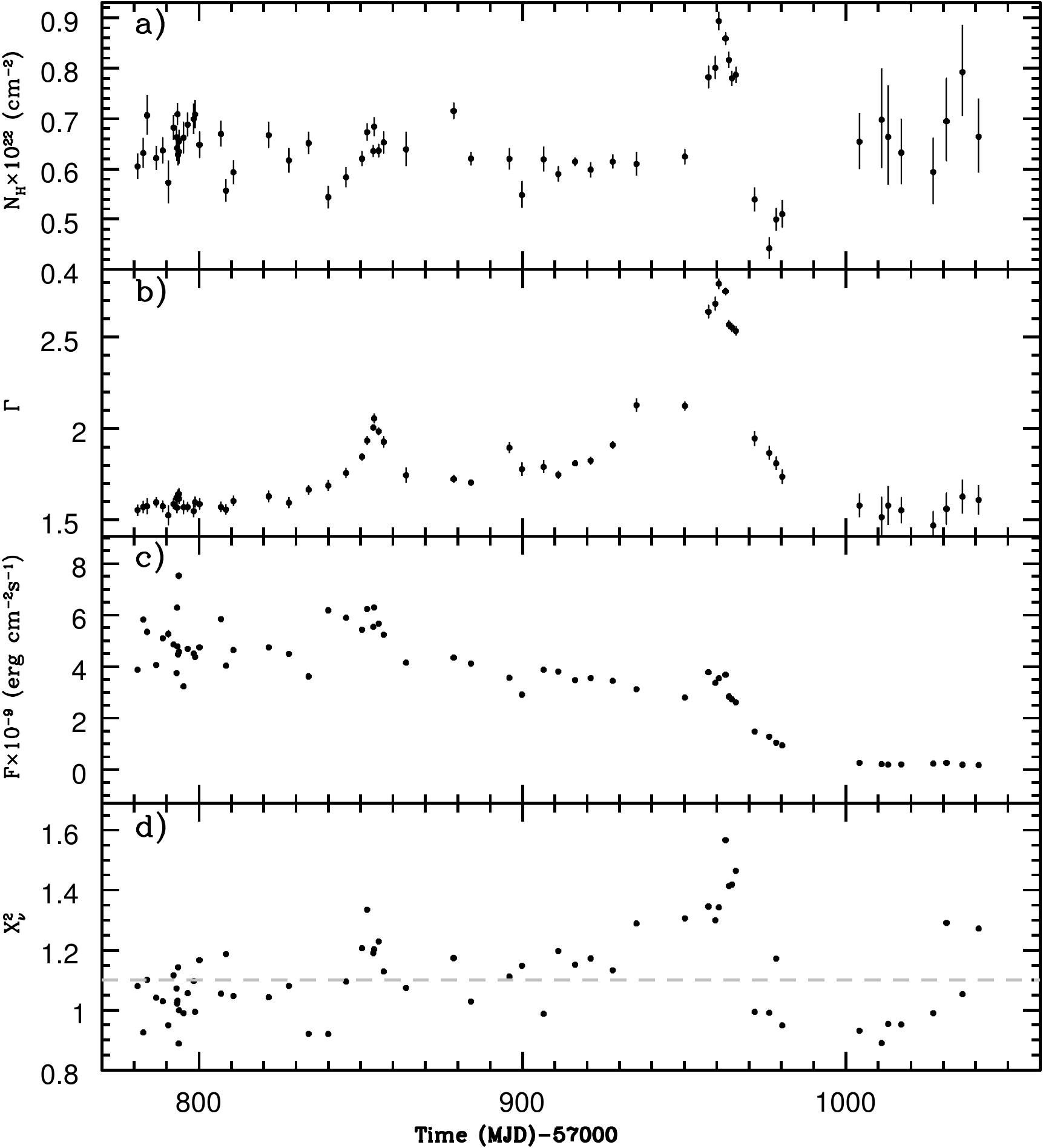}
   \caption{Parameters of the XRT spectra fitted with an absorbed power law model ({\sc tbabs*po}). From the top to the bottom panel: a) hydrogen column density in units of $10^{22}$\,cm$^{-2}$, b) photon index, c) 0.5--10\,keV absorbed flux in units of $10^{-9}$ \ergcm, d) reduced $\chi^{2}$. 
 The high variation of N$_{\rm H}$ is due to a non adequate spectral modeling. The spectra with a \chir > 1.1 and $\Gamma\geq$ 1.8 require an additional multicolor disc black-body ({\sc diskbb}) component.}
              \label{fit_xrt}
\end{figure*}

We fit each XRT spectrum in the 0.5--10\,keV energy band with an absorbed ({\sc tbabs} in XSPEC) power-law model with \cite{wilms00} cosmic abundances and \cite{verner96} cross-sections for the interstellar absorption. In Figure \ref{fit_xrt} we show the evolution of the spectral parameters.   
N$_{\rm H}$ shows values around (0.6-0.7) $\times$ 10$^{22}$\,\cm, with a peak value at $\sim$0.9 $\times$ 10$^{22}$\,\cm~ (MJD 57960.72). 
The photon index ($\Gamma$) remains between 1.5 and 1.7 throughout the whole outburst, with the exception of significant increases during the three XRT peaks and BAT dips observed over the periods of spectral softening (Fig. \ref{lc_hr}). The XRT spectrum corresponding to the first peak (2017 April 11, $\#$032) shows a photon index $\Gamma\sim$2.1, while on May 22 ($\#$038) we obtained a photon index of $\sim$1.9. The source reached its steepest spectral slope on July 26 ($\#$050) with $\Gamma\sim$2.8, corresponding to the lowest HR value and the highest N$_{\rm H}\sim$0.9 $\times$ $10^{22}$\,\cm, even though the reduced $\chi^{2}$ is not statistically acceptable (see Tab. \ref{log}). 
The large column density variations derived could be caused by the inadequacy of our model, i.e. an accretion disc could be required to fit the data.
We therefore added a multicolor disc black-body ({\sc diskbb} in XSPEC) component to the previous model in all spectra with \chir\ > 1.1. We estimated with the F-test that this component is required by all spectra with $\Gamma\geq$ 1.8.
However, we noted that in a few spectra where $\Gamma$ < 1.8 with \chir\ > 1.1, the bad residuals are due to the known strong instrumental silicon (1.84\,keV) and gold (2.2\,keV) edges\footnote{\url{https://heasarc.gsfc.nasa.gov/docs/heasarc/caldb/swift/docs/xrt/SWIFT-XRT-CALDB-09\_v19.pdf}}, so these fits can be considered acceptable.\\

\subsection{Broadband X-ray spectra} 
\label{broadband}
\begin{figure*}
   \centering
    \includegraphics[width=15cm]{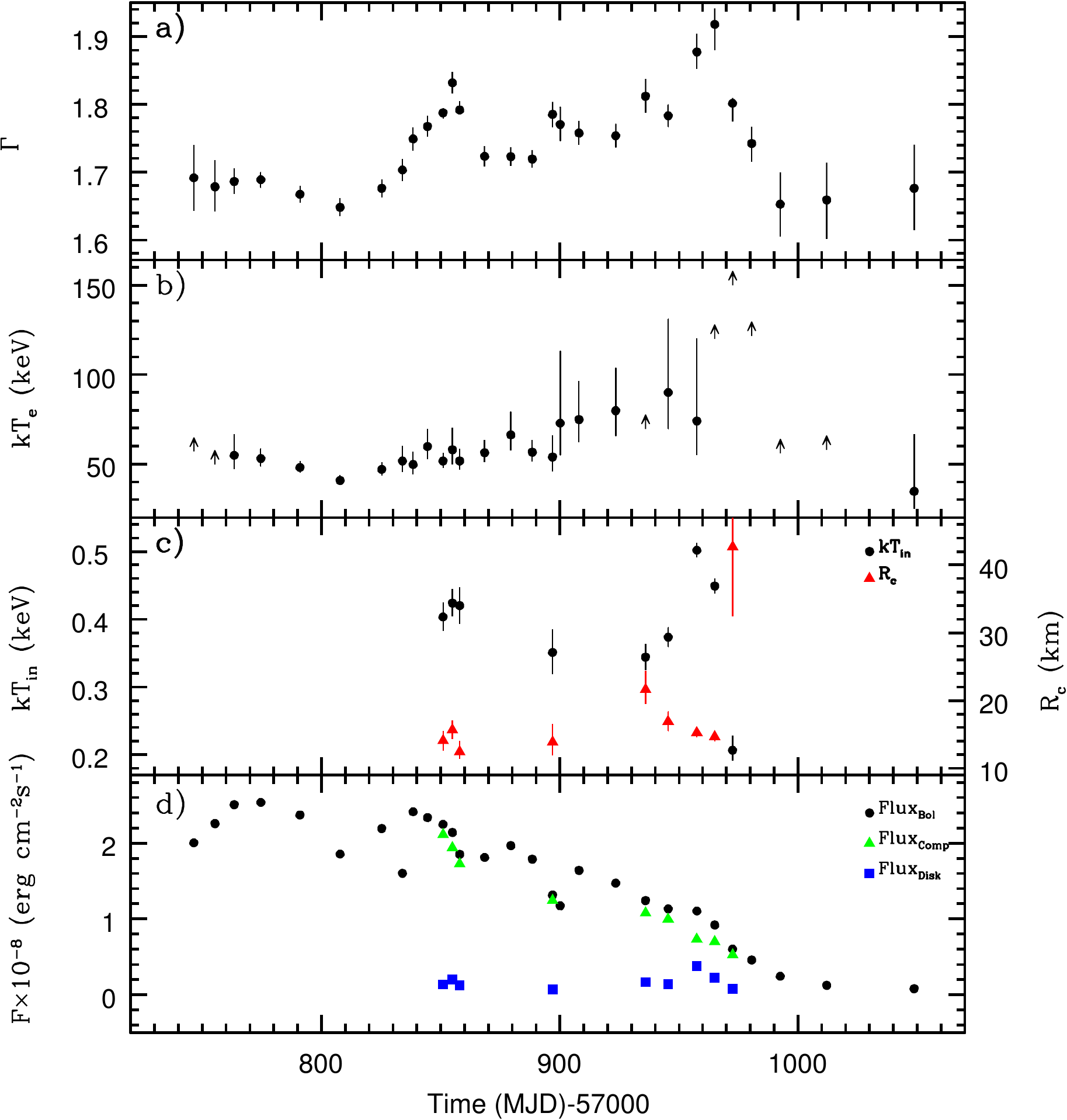}

   \caption{Evolution of the main spectral parameters and flux obtained fitting the broadband spectra with an absorbed thermal Comptonisation model ({\sc tbabs*nthcomp}) plus a disc black-body component  ({\sc diskbb}) when required (parameters reported in Tab. \ref{gti_broad}).
From the top to the bottom panel: a) photon index, b) comptonising electron temperature in units of keV, c) inner disc black-body temperature in units of keV (black dots) and inner disc radius depending on the inclination angle (R$_{\rm c}$) in units of km (red triangles) and d) 0.1--500\,keV unabsorbed bolometric flux in units of $10^{-8}$\,\ergcm (black dots). The green triangles show the Comptonised flux and the blue squares show the disc flux. The disc luminosity never dominates the energetic of the source.}
              \label{fit_broadband}
\end{figure*}

\begin{figure*}
   \centering
    \includegraphics[angle=-90,width=8cm]{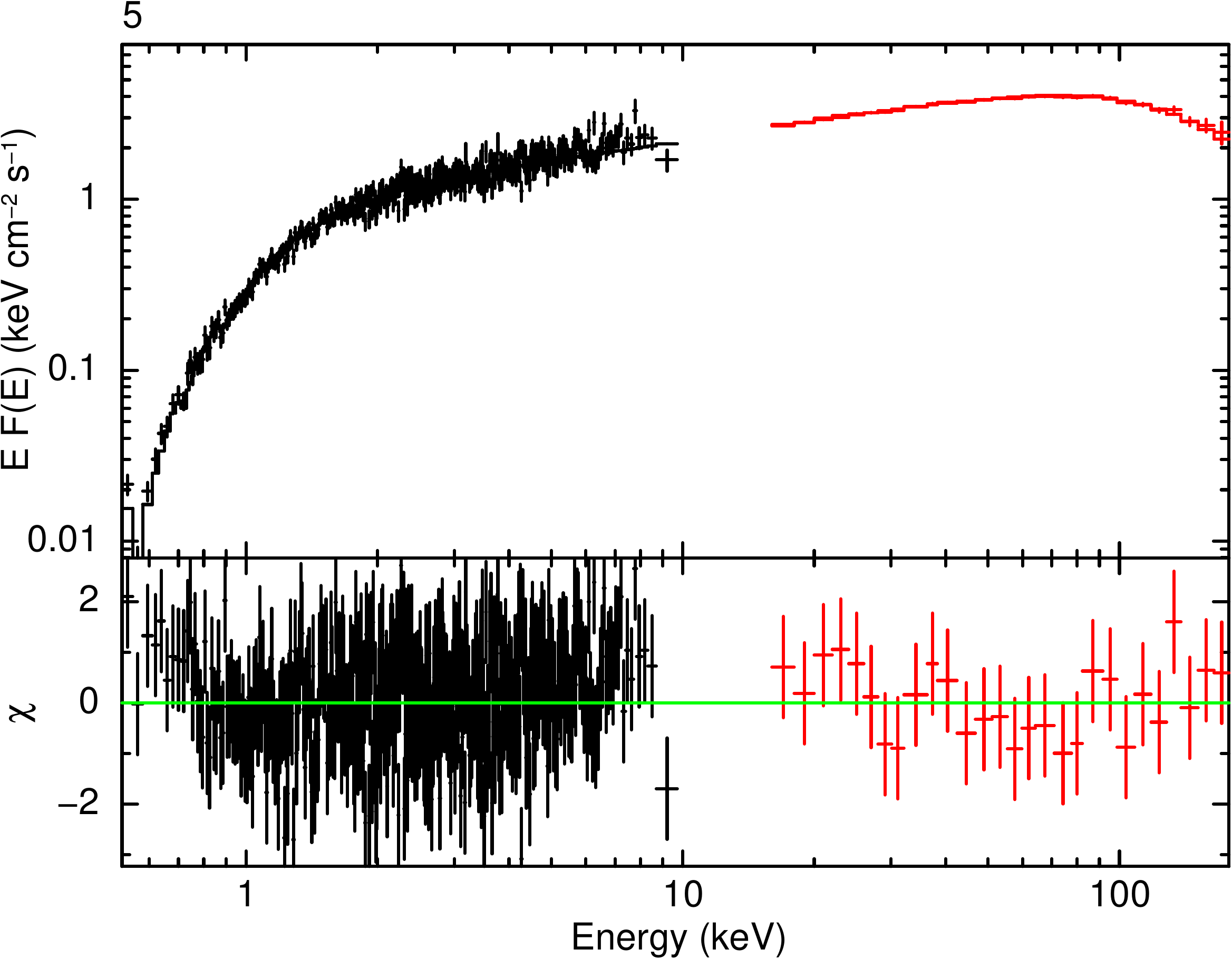}
    \includegraphics[angle=-90,width=8cm]{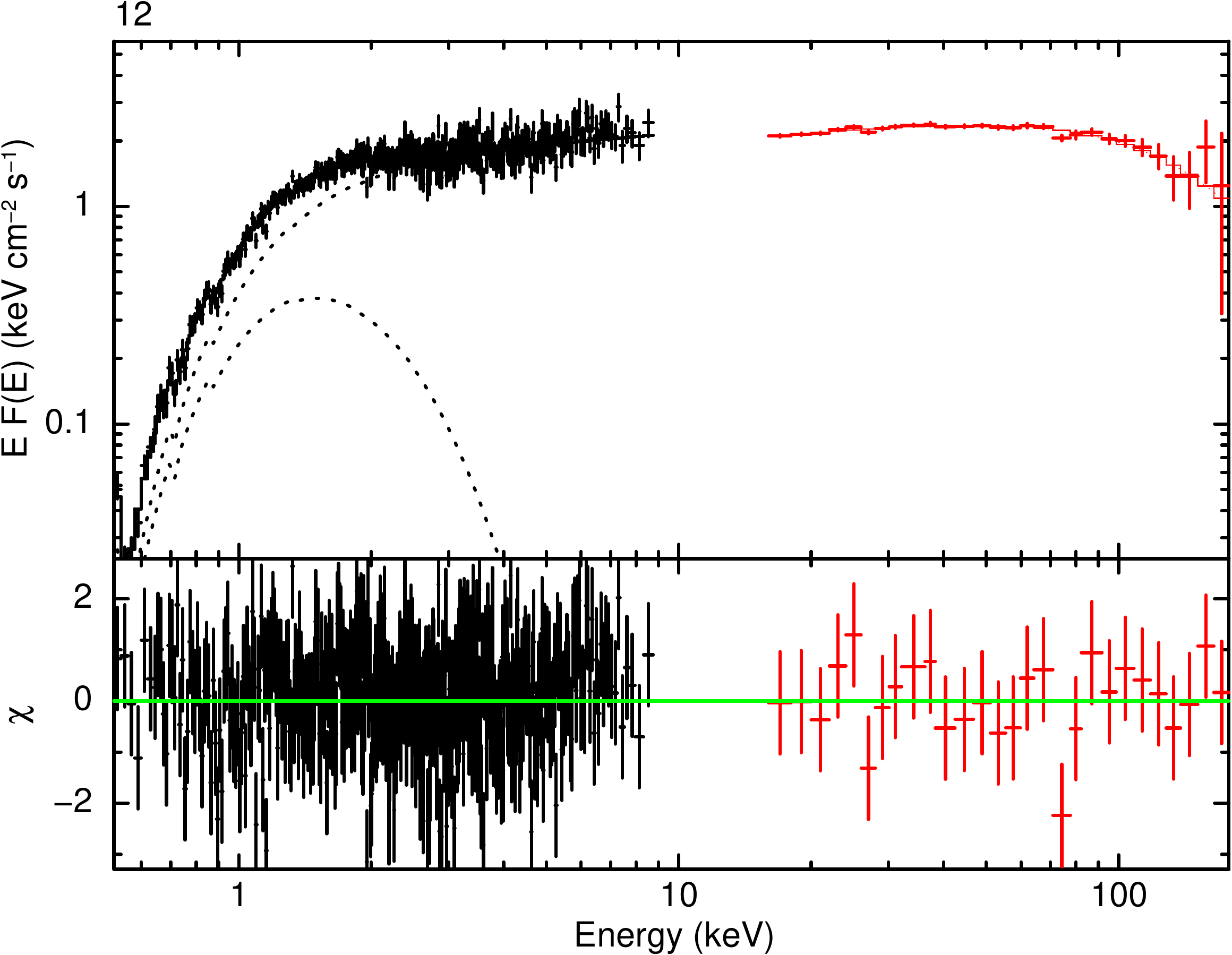}
    \\
    \includegraphics[angle=-90,width=8cm]{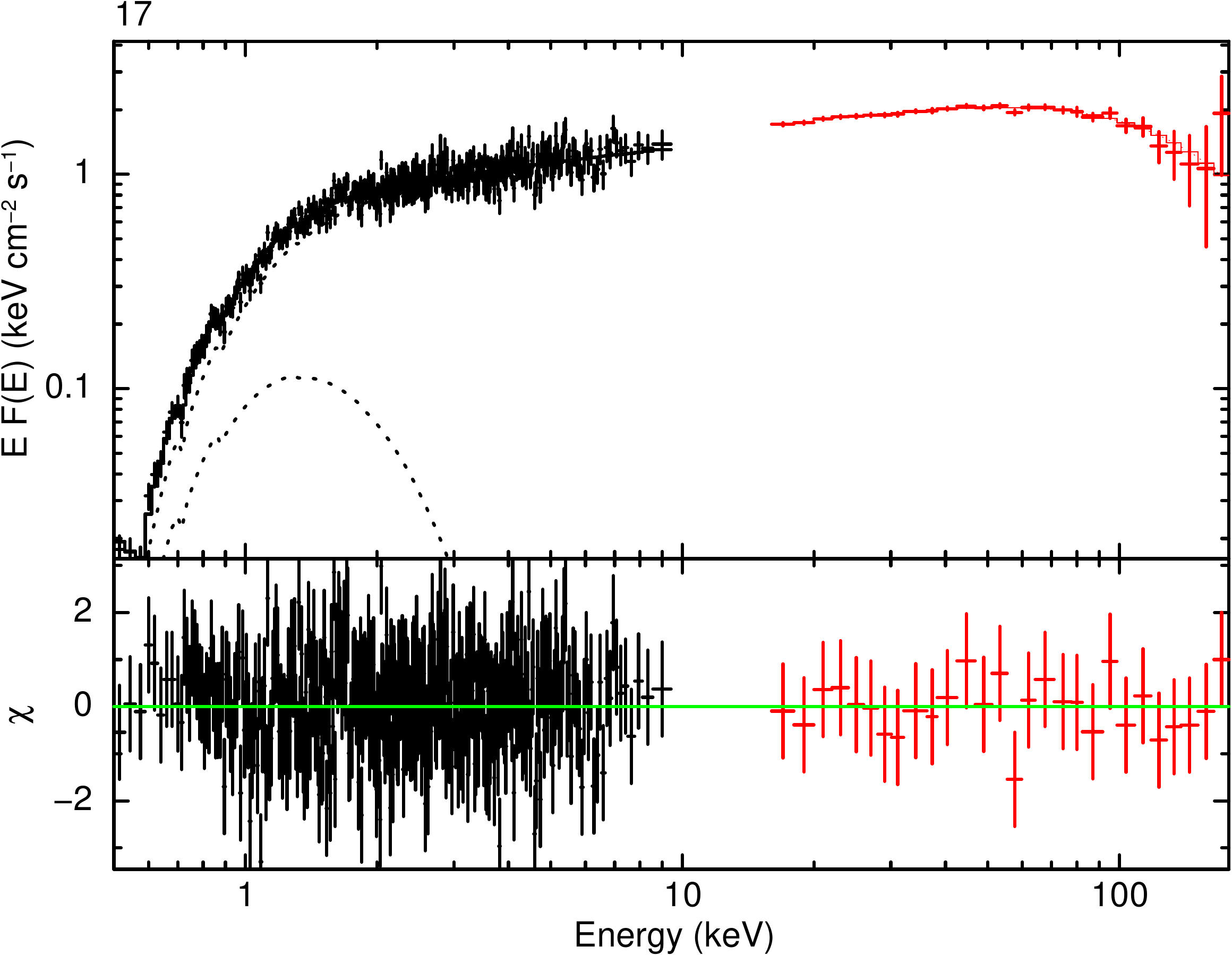}
    \includegraphics[angle=-90,width=8cm]{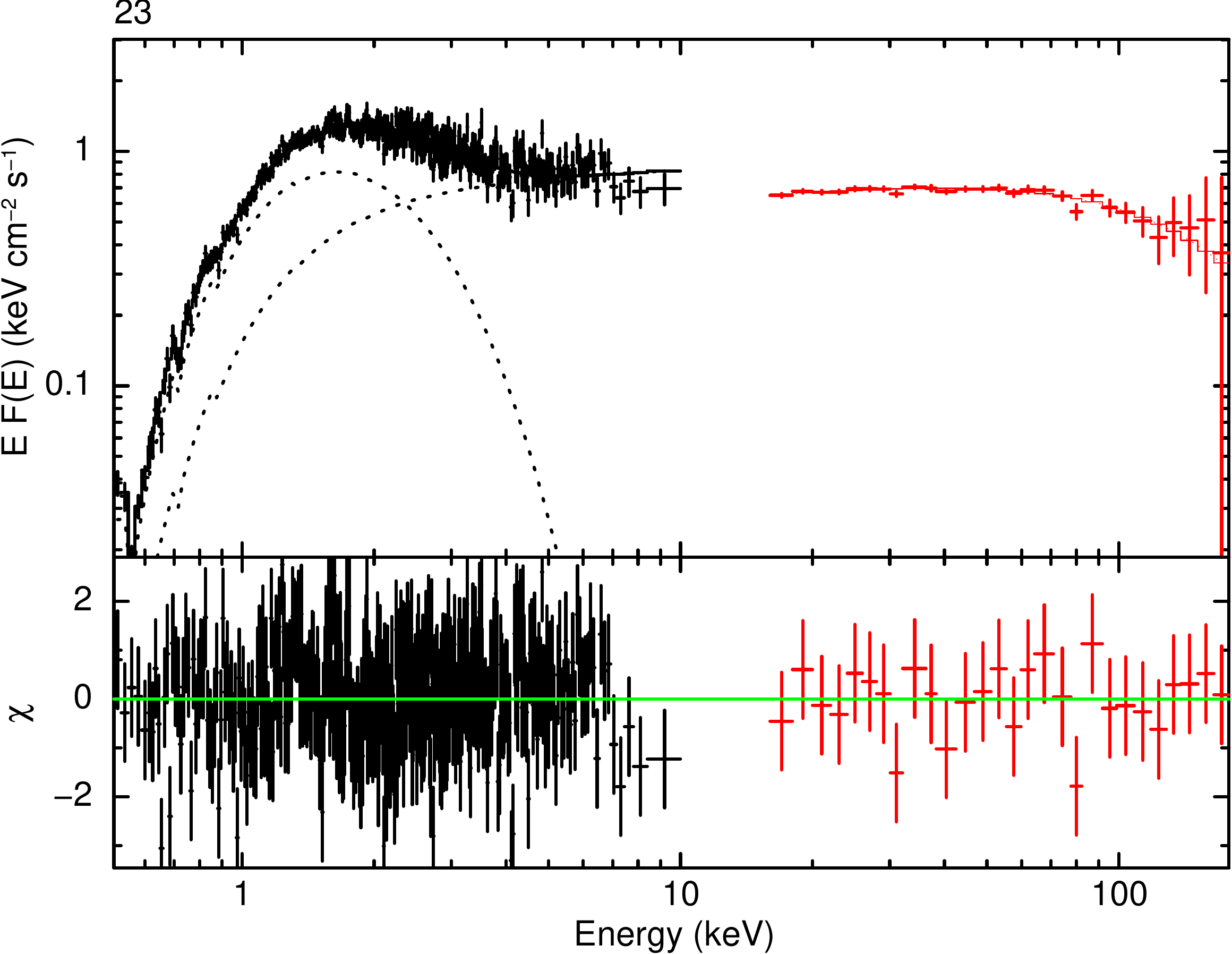}
    \\
    \includegraphics[angle=-90,width=8cm]{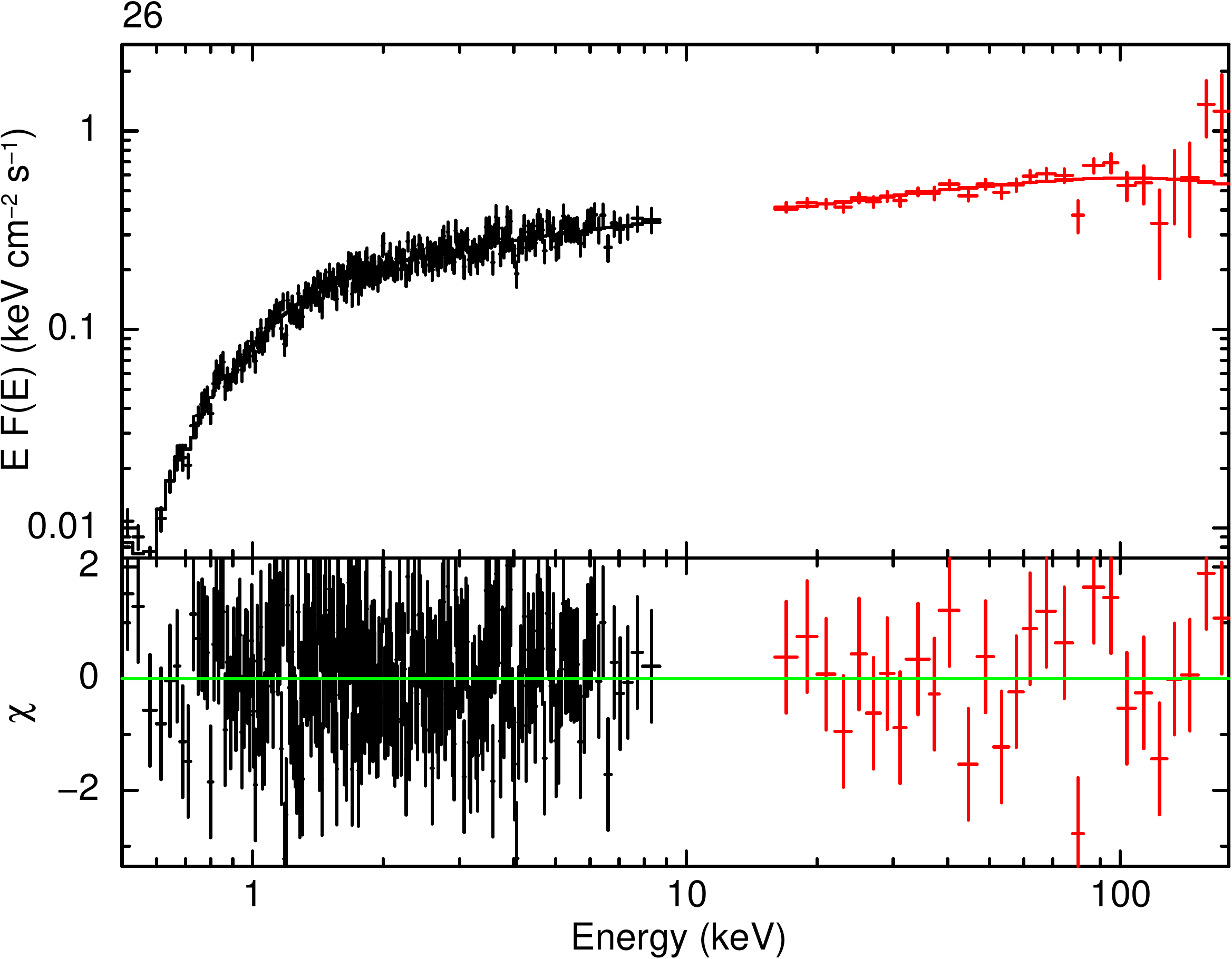}
    \includegraphics[angle=-90,width=8cm]{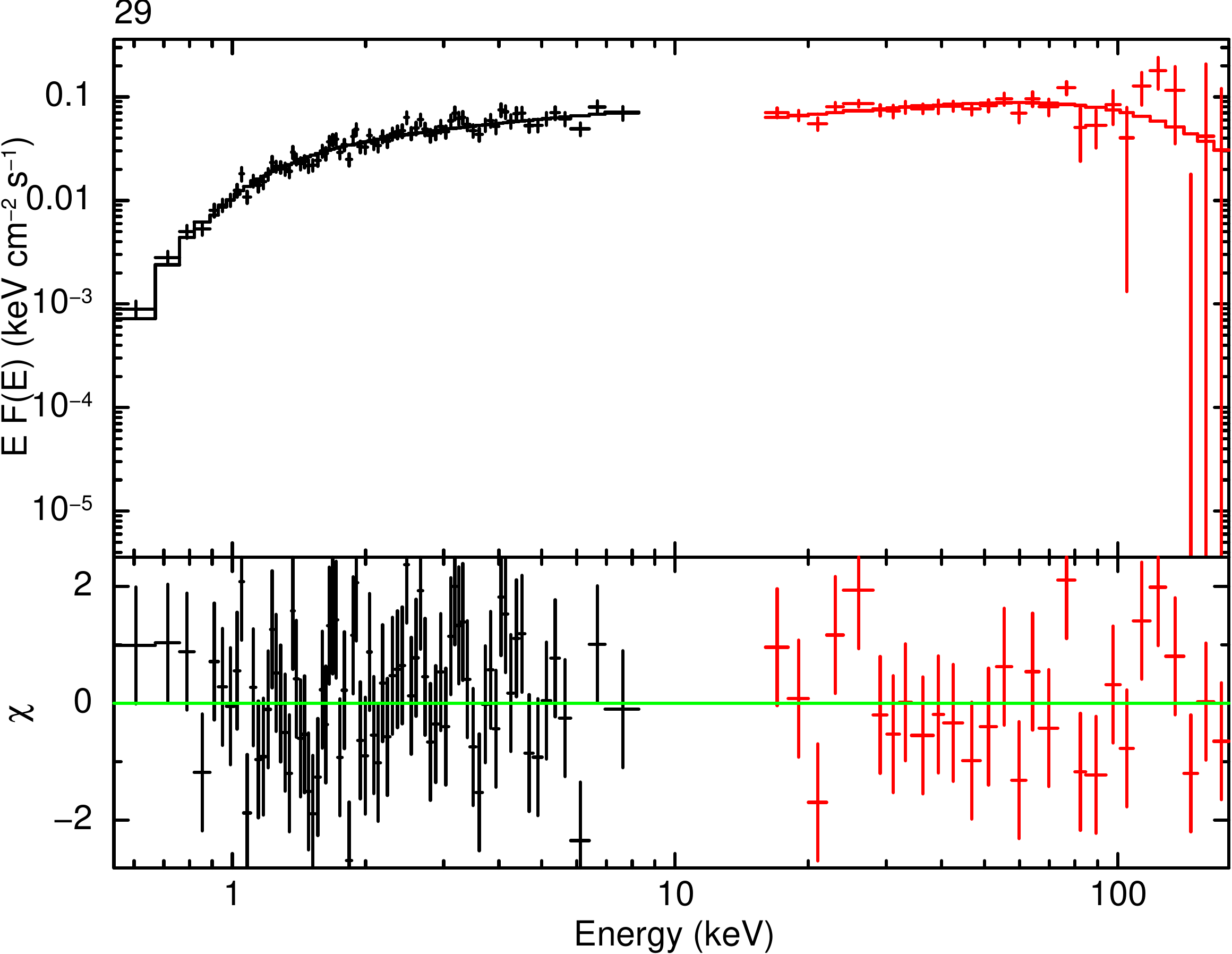}

   \caption{Broadband energy spectra of six different XRT pointings ($\#5$, $\#32$, $\#38$, $\#50$, $\#57$ and $\#66$, see Tab. \ref{log}) with average BAT spectra in the corresponding quasi-simultaneous GTIs. They are fitted with an absorbed Thermal Comptonisation model ({\sc tbabs*nthcomp}) plus a multicolour disc black-body component ({\sc diskbb}) when required ({\it 12}, {\it 17} and {\it 23}). The spectral parameters are reported in Tab. \ref{results} (spectra {\it 5}, {\it 12}, {\it 17}, {\it 23}, {\it 26} and {\it 29}).}
              \label{fit_spectra}
\end{figure*}

We improved the spectral analysis by adding averaged BAT spectra in order to perform a broadband fit. 
Using the light curves and HRs (see Fig. \ref{lc_hr}), we selected good time intervals (GTIs, see Tab. \ref{gti_broad}) based on two selection criteria: 1) constant HR in both BAT and XRT, 2) BAT flux variation lower than 20$\%$, with the exception of the two last intervals where the variation is about 30$\%$. We point out that there are a number of BAT GTIs averaged spectra which do not overlap with any XRT observations.    
On the other hand, we selected one XRT spectrum for each GTI, making sure that it was consistent, in terms of spectral parameters and flux, with the other XRT spectra within the same GTI.  
In the broadband spectra we introduced a systematic error of 2$\%$ according to the instrument calibration guidelines.\\
The broadband spectra were fitted with an absorbed thermal Comptonisation model ({\sc tbabs} plus {\sc nthcomp} in XSPEC, see Tab. \ref{gti_broad}). We assumed that the seed photons followed a disc black-body distribution (parameter {\it inp$\_$type}$=$1) and we fixed the seed photon temperature at 0.1\,keV every time the disc black-body component was not required.
We verified that the {\sc diskbb} component, when required, improved our fits significantly (F-test values always lower than 8$\times$10$^{-5}$).
When fitting all spectra with a constant N$_{\rm H}$, we noticed statistically significant discrepancies in the residuals of several spectra, so that we left this parameter free to vary.
However, we fitted simultaneously the nine spectra in which the disc component was required, keeping the N$_{\rm H}$ tied (see Tab. \ref{gti_broad}).\\
The photon index varies from 1.6 up to 1.9 (see panel a in Fig. \ref{fit_broadband}) while the electron temperature (\kte, see panel b in Fig. \ref{fit_broadband}) is well constrained when the source is bright in hard X-ray.  However, in some cases, in particular during the last spectral softening, we only obtained a lower limit on this parameter. Extra components in addition to the thermal Comptonisation could play a role  (i.e., non-thermal Comptonisation, reflection) in the intermediate states.
Nevertheless, the BAT statistics and the poor high energy coverage do not allow us to disentangle the X-ray continuum of BHTs.\\
The thermal disc component is well detected in the spectra collected during the three peaks observed in the XRT light curve.
The inner disc temperature (\kti) varies between 0.2 and 0.5\,keV, the latter was observed during the third peak softening (see panel c in Fig. \ref{fit_broadband}). Our findings are in agreement with \citet{munoz11} who found that the presence of the accretion disc is detected when the fractional \rms\ is $\sim$20\%.
In the spectra collected during the peak of the three softenings (spectra {\it 12}, {\it 17} and {\it 23} in Tab. \ref{gti_broad}), we estimated that the disc flux contributes 9$\%$, 5$\%$ and 34$\%$, respectively, to the total unabsorbed bolometric flux of \grs. 
In Fig. \ref{fit_spectra} we show energy spectra, models and residuals for six representative observations: the three softening peaks (HIMS), one bright HS spectrum ({\it 5}), one HIMS spectrum collected during the outburst decreasing phase ({\it 26}) and a HS spectrum ({\it 28}) collected before the source quiescence.

\begin{figure*}
   \centering
   \includegraphics[angle=-90, width=7cm]{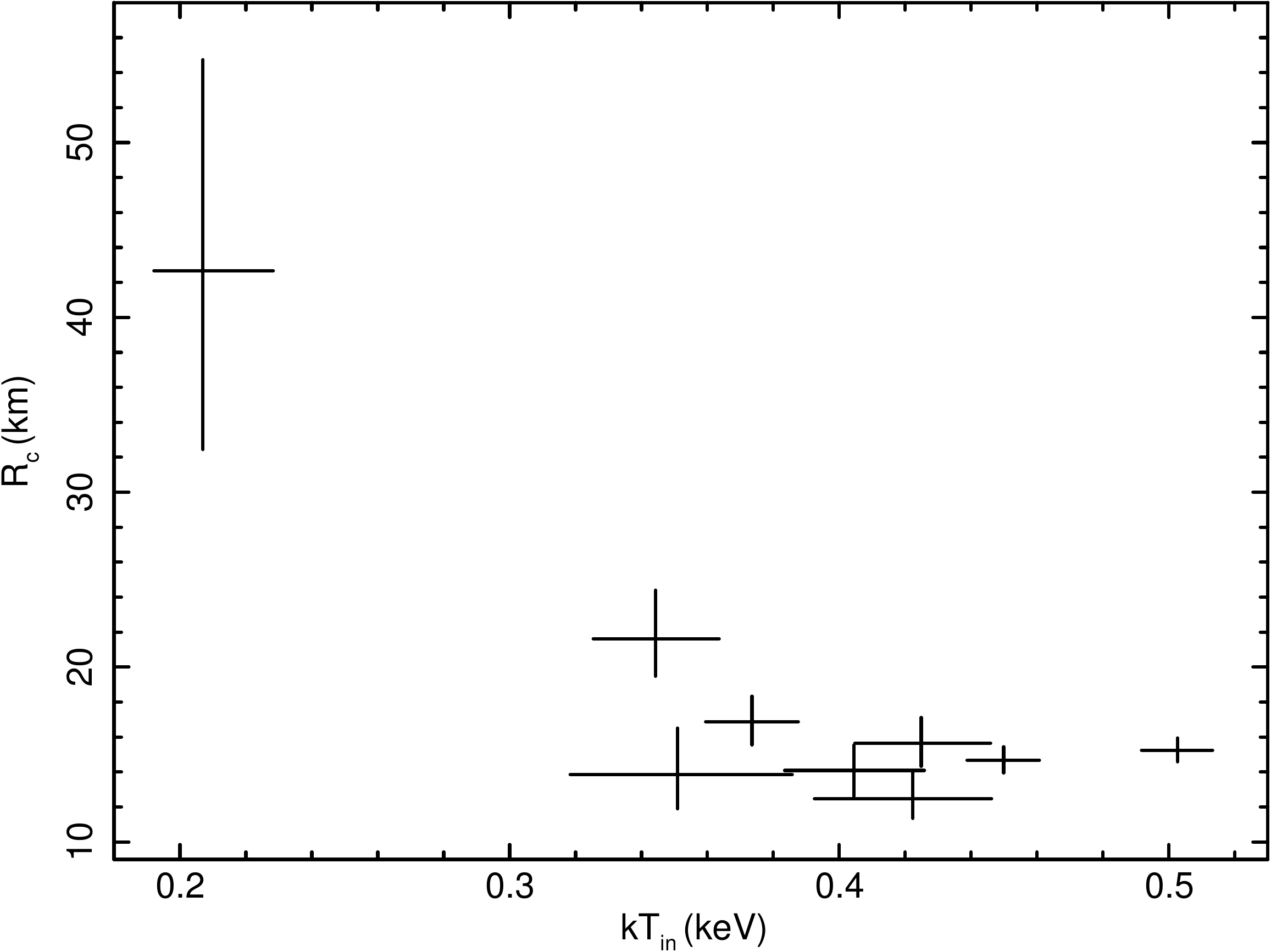}
    \includegraphics[angle=-90, width=7cm]{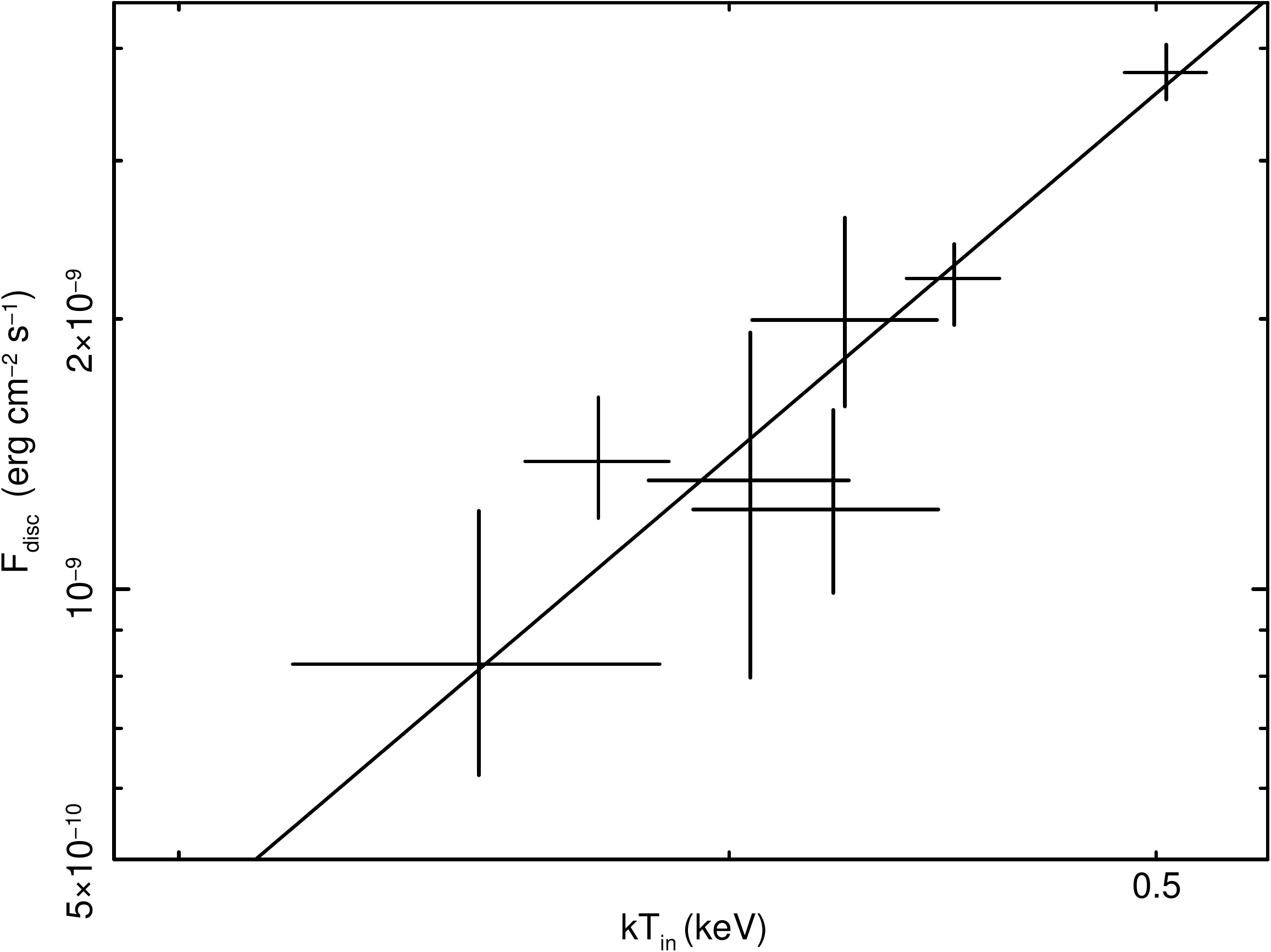}
   \caption{{\it Left panel:} Inner disc radius depending on the inclination angle (R$_{\rm c}$) in function of the inner disc temperature. R$_{\rm c}$ decreases with the increase of the temperature until it reaches an almost constant value. {\it Right panel:} the observed disc flux versus the inner disc temperature including only spectra with R$_{\rm c}$ almost constant. The slope of the power-law in log-log is $\alpha= 4.2 \pm 1.0$, in agreement with a constant inner disc radius scenario. }
        \label{fig:kt_fl}
\end{figure*}

In addition, we have estimated the inner disc radius as a function of the inclination angle (see Tab. \ref{gti_broad}) from the {\sc diskbb} normalization. Then we applied the correction factor between the apparent and true inner radius by \cite{kubota98} and the hardening factor of 1.7 \citep{shimura95}.
In Fig. \ref{fig:kt_fl} (left), we plot R$_{\rm in}$ (cos$\vartheta)^{1/2}$ (hereafter R$_{\rm c}$) versus \kti\ which shows that most of our measurements are consistent with a constant radius R$_{\rm c} \sim$15\,km. There is only one point (at lower temperature) that appears to have a significantly higher inner disc radius and a second point with a value ($\sim$23\,km) which is not consistent with the mean radius. 
We also estimated whether the disc luminosity varies as a function of the inner disc temperature according to L $\propto$ T$^{4}$.
In Fig. \ref{fig:kt_fl} (right) we plot the observed disc flux and temperature against each other for all observations where R$_{\rm c}$ was almost constant.
The power-law index from the best-fit is $\alpha= 4.2 \pm 1.0$ ($\chi^{2}$(dof) = 4.5(5)). The probability to have the $\chi^{2}$ obtained is of 0.52$\%$.
This suggests that the inner disc temperature variations are driven only by changes in mass accretion rate at constant inner disc radius, and appears inconsistent with a varying inner disc radius at constant mass accretion rate \citep[L $\propto $ T$^{4/3}$;][]{done07}.
This is particularly evident within the third softening when a flux variation of about a factor of three is observed in combination with a constant R$_{\rm c}$ and an increase of \kti\ (see Tab. \ref{gti_broad} and Fig. \ref{fit_broadband}, c).

\subsection{Radio and correlation with X-rays} 
\label{radio}
\begin{figure*}
   \centering
    \includegraphics[width=15cm]{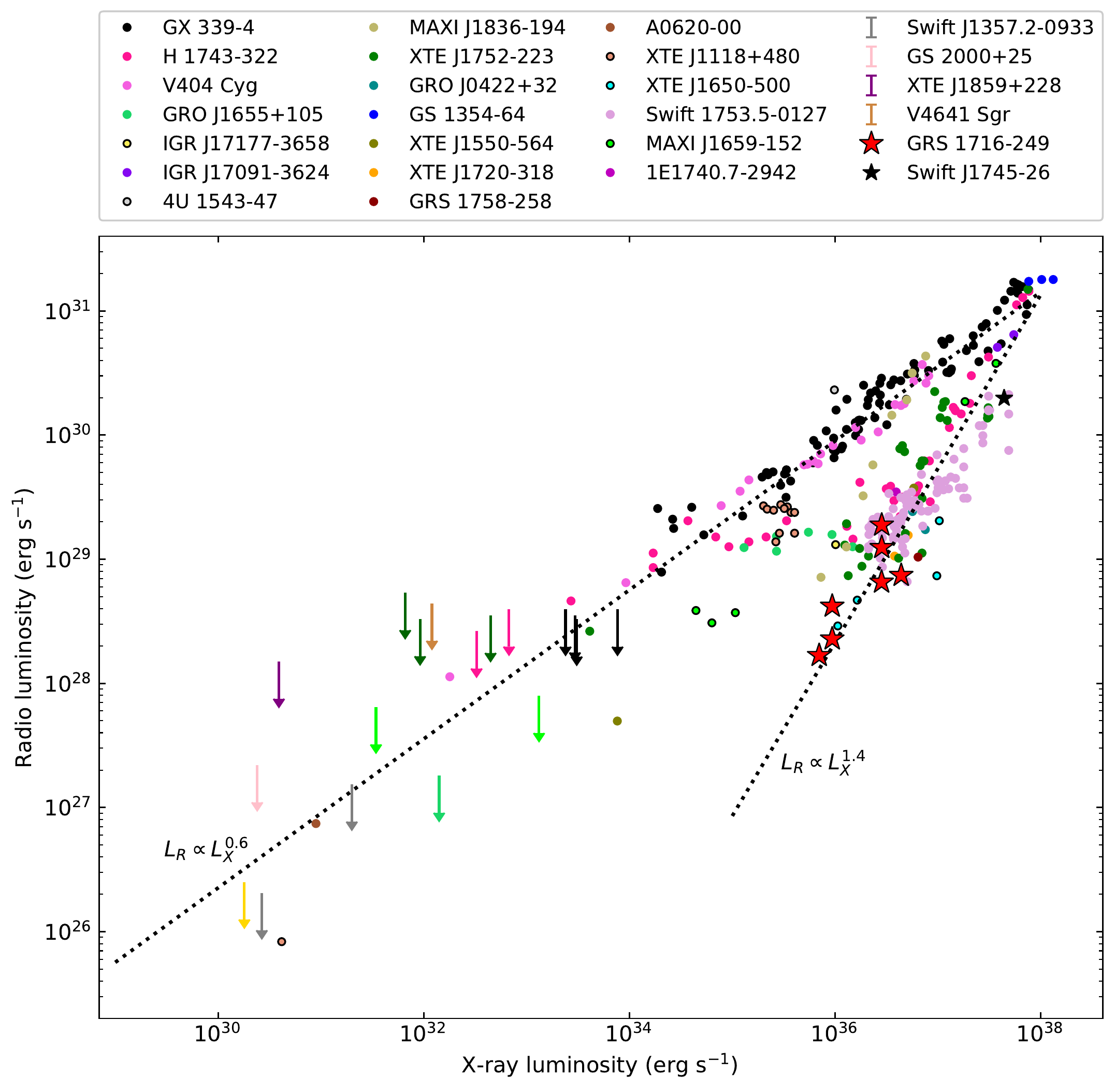}

   \caption{Radio/X-ray luminosity correlation. The X-ray luminosities have been calculated in the 1--10\,keV energy range and the radio observations, collected at different frequencies, have been converted by \citealt{baharaian18} to a common radio frequency, i.e. $\sim$5\,GHz. \grs~ (red stars) is located on the radio-quiet branch. For this source we plotted radio observations performed both at $\sim$5 and $\sim$9\,GHz.}
              \label{x-radio}
\end{figure*}

In Figure \ref{x-radio}, we show the radio/X-ray luminosity correlation combining data made available by \cite{baharaian18} and references therein\footnote{\url{https://github.com/arushton/XRB-LrLx_pub}}. All X-ray luminosities are calculated in the 1--10\,keV energy range.
We also added a simultaneous XRT (in hard state, see \citealt{delsanto16}) and radio observation performed on MJD 56187.99 of the source Swift J174510.8--262411 (hereafter \swf) which has been never taken into account in previous radio/X-ray luminosity correlation planes. 
We estimated a X-ray luminosity for \swf~ of 4.4 $\times$ 10$^{37}$\,\ergs~ at a distance of 7\,kpc \citep{munozdarias13} and we used the 5\,GHz radio flux density from \cite{curran14}.\\
We measured the X-ray flux of \grs~in the 1--10\,keV energy band, using data that were quasi-simultaneous with the radio observations performed at $\sim$5 and $\sim$9\,GHz (see Tab. \ref{tab:radio}). We estimated X-ray luminosities (d=2.4\,kpc) of 2.8 $\times$ 10$^{36}$\,\ergs~on MJD 57793, 4.4 $\times$ 10$^{36}$\,\ergs~on MJD 57805, 2.8 $\times$ 10$^{36}$\,\ergs~on MJD 57865, 9.4 $\times$ 10$^{35}$\,\ergs~on MJD 57977 and 7.0 $\times$ 10$^{35}$\,\ergs~on MJD 57978. 
\grs~ is located on the branch consistent with the slope of 1.4.\\
The radio spectral indices measured by ATCA at the beginning ($\alpha=$ -0.15$\pm$0.08 on MJD 57793.835) and close to the end of the outburst ($\alpha = $ -0.07$\pm$0.19 on MJD 57977.256) by VLA are both consistent with a flat-spectrum compact jet.
\cite{espinasse18} performed a study on a sample of 17 bright BHBs in hard state. They show that the radio spectral indices distribution could be related to the position of the sources on the radio/X-ray luminosity correlation plane (see Sec. \ref{discussion}).

\vfill
\begin{landscape}
\begin{table}
\begin{center}
\caption{Good Time Intervals and best-fit parameters of the broadband XRT and BAT spectra performed with an absorbed thermal Comptonisation ({\sc nthcomp}) model and a disc black-body model ({\sc diskbb}) when required. (1) Number of the spectrum, (2) sequence number of the XRT pointing selected within the BAT GTI, start (3) and stop (4) time of the GTI in MJD, (5) hydrogen column density in units of $10^{22}$\,cm$^{-2}$, (6) photon index, (7) comptonising electron temperature in keV, (8) inner disc temperature, (9) inner disc radius, (10) unabsorbed bolometric flux (0.1--500\,keV), disc  and Comptonised component fluxes, (11) reduced $\chi^{2}$.\newline
$^{\ast}$ broadband spectra fitted all together keeping linked N$_{\rm H}$ and letting the other parameters to vary.}   
\label{gti_broad}     
\centering     
\small 
{\renewcommand\arraystretch{1.2}    
\begin{tabular}{lcccccccccccc}
\hline\hline               
n&Seq.  & MJD Start & MJD Stop & N$_{\rm H}\times$ 10$^{22}$ & $\Gamma$ & \kte & \kti &R$_{\rm in}$
 (cos$\vartheta)^{1/2}$& & Flux $\times$ 10$^{-8}$ &  & $\chi^{2}_{r}$(dof) \\    
 &   ($\#$)   &  &          &(cm$^{-2}$)&        & (keV)    & (keV)             & (km)      &      & (\ergcm)                                      &    &                 \\
 &    &     &          &           &        &          &          &    &       Bol & Disc & Comp & \\
(1)&  (2) & (3) & (4) & (5) & (6)& (7)& (8)&(9)& &(10)&&(11)\tabularnewline
\hline           

\\
{\it 1}  & -   & 57740.08 & 57752.85 & -                      & 1.69$^{+0.05}_{-0.05}$ & >57               & -                        &-& 2.01 & -     &   2.01  & 1.22(25)  \tabularnewline 
{\it 2}  & -   & 57753.24 & 57757.49 & -                      & 1.68$^{+0.04}_{-0.04}$ & >50               & -                        &-& 2.26 & -     &   2.26  & 1.01(25)  \tabularnewline 
{\it 3}  & -   & 57760.41 & 57766.39 & -                      & 1.69$^{+0.02}_{-0.02}$ & 55$^{+12}_{-8}$   & -                        &-& 2.51 & -     &   2.51  & 0.91(25)  \tabularnewline 
{\it 4}  & -   & 57769.24 & 57779.96 & -                      & 1.69$^{+0.01}_{-0.01}$ & 53$^{+5}_{-4}$    & -                        &-& 2.54 & -     &   2.54  & 0.97(25)  \tabularnewline 
{\it 5}  & 005 & 57780.01 & 57801.96 & 0.69$^{+0.02}_{-0.02}$ & 1.67$^{+0.01}_{-0.01}$ & 48$^{+4}_{-3}$    & -                        &-& 2.37 & -     &   2.37  & 1.01(399) \tabularnewline 
{\it 6}  & 020 & 57802.01 & 57813.70 & 0.61$^{+0.02}_{-0.02}$ & 1.65$^{+0.01}_{-0.01}$ & 41$^{+2}_{-3}$    & -                        &-& 1.86 & -     &   1.86  & 1.18(388) \tabularnewline 
{\it 7}  & 022 & 57818.61 & 57831.99 & 0.70$^{+0.02}_{-0.02}$ & 1.68$^{+0.01}_{-0.01}$ & 47$^{+4}_{-3}$    & -                        &-& 2.19 & -     &   2.19  & 0.99(401) \tabularnewline 
{\it 8}  & 025 & 57832.02 & 57835.98 & 0.67$^{+0.02}_{-0.02}$ & 1.70$^{+0.02}_{-0.02}$ & 52$^{+9}_{-6}$    & -                        &-& 1.60 & -     &   1.60  & 0.89(410) \tabularnewline 
{\it 9}  & 026 & 57836.02 & 57840.96 & 0.57$^{+0.02}_{-0.02}$ & 1.75$^{+0.02}_{-0.02}$ & 50$^{+7}_{-5}$    & -                        &-& 2.42 & -     &   2.42  & 0.90(359) \tabularnewline 
{\it 10} & 027 & 57841.00 & 57848.00 & 0.58$^{+0.02}_{-0.02}$ & 1.77$^{+0.02}_{-0.02}$ & 60$^{+10}_{-7}$   & -                        &-& 2.34 & -     &   2.34  & 1.03(415) \tabularnewline 

{\it 11}$^{\ast}$ & 030 & 57848.06 & 57853.97 & 0.59$^{+0.01}_{-0.01}$ & 1.79$^{+0.01}_{-0.01}$ & 52$^{+5}_{-4}$    & 0.40$^{+0.02}_{-0.02}$ & 14.1$^{+1.5}_{-1.5}$  & 2.25 & 0.13  &  2.12   & 1.12(3411) \tabularnewline 
{\it 12}$^{\ast}$ & 032 & 57854.02 & 57855.96 & 0.59$^{+0.01}_{-0.01}$ & 1.83$^{+0.02}_{-0.02}$ & 58$^{+13}_{-8}$   & 0.42$^{+0.02}_{-0.02}$ & 15.6$^{+1.5}_{-1.3}$  & 2.14 & 0.20  &  1.94   & 1.12(3411) \tabularnewline 
{\it 13}$^{\ast}$ & 034 & 57856.01 & 57860.00 & 0.59$^{+0.01}_{-0.01}$ & 1.79$^{+0.01}_{-0.01}$ & 52$^{+7}_{-5}$    & 0.42$^{+0.03}_{-0.03}$ & 12.5$^{+1.6}_{-1.1}$  & 1.85 & 0.12  &  1.73   & 1.12(3411) \tabularnewline 

{\it 14} & 035 & 57860.00 & 57876.96 & 0.62$^{+0.02}_{-0.02}$ & 1.72$^{+0.02}_{-0.02}$ & 56$^{+7}_{-5}$    & -                        &- & 1.81 & -     &   1.81  & 1.00(262) \tabularnewline 
{\it 15} & 036 & 57877.00 & 57881.95 & 0.71$^{+0.01}_{-0.01}$ & 1.72$^{+0.01}_{-0.01}$ & 66$^{+13}_{-9}$   & -                        &- & 1.97 & -     &   1.97  & 1.09(550) \tabularnewline 
{\it 16} & 037 & 57882.00 & 57894.97 & 0.62$^{+0.01}_{-0.01}$ & 1.72$^{+0.01}_{-0.01}$ & 57$^{+7}_{-5}$    & -                        &- & 1.79 & -     &   1.79  & 0.96(563) \tabularnewline 

{\it 17}$^{\ast}$ & 038 & 57895.02 & 57898.94 & 0.59$^{+0.01}_{-0.01}$ & 1.79$^{+0.02}_{-0.02}$ & 54$^{+12}_{-8}$   & 0.35$^{+0.03}_{-0.03}$ &  13.9$^{+2.7}_{-2.0}$  & 1.32 & 0.07  &  1.24   & 1.12(3411) \tabularnewline 

{\it 18} & 039 & 57899.05 & 57901.32 & 0.54$^{+0.02}_{-0.02}$ & 1.77$^{+0.03}_{-0.02}$ & 73$^{+40}_{-18}$  & -                       &-& 1.17 & -     &   1.17  & 1.11(287) \tabularnewline 
{\it 19} & 040 & 57902.27 & 57913.89 & 0.59$^{+0.02}_{-0.02}$ & 1.76$^{+0.02}_{-0.02}$ & 75$^{+21}_{-13}$  & -                       &-& 1.64 & -     &   1.64  & 0.93(321) \tabularnewline 
{\it 20} & 043 & 57914.01 & 57932.96 & 0.54$^{+0.01}_{-0.01}$ & 1.75$^{+0.02}_{-0.02}$ & 80$^{+24}_{-14}$  & -                       &-& 1.47 & -     &   1.47  & 0.96(380) \tabularnewline 

{\it 21}$^{\ast}$ & 045 & 57933.01 & 57938.95 & 0.59$^{+0.01}_{-0.01}$ & 1.81$^{+0.02}_{-0.02}$ & >70               & 0.34$^{+0.01}_{-0.01}$ & 21.6$^{+2.8}_{-2.1}$  & 1.24 & 0.16  & 1.08    & 1.12(3411) \tabularnewline 
{\it 22}$^{\ast}$ & 047 & 57939.05 & 57951.97 & 0.59$^{+0.01}_{-0.01}$ & 1.78$^{+0.02}_{-0.02}$ & 90$^{+41}_{-21}$  & 0.37$^{+0.01}_{-0.01}$ & 16.9$^{+1.4}_{-1.3}$  & 1.13 & 0.14  & 0.99    & 1.12(3411) \tabularnewline 
{\it 23}$^{\ast}$ & 050 & 57952.09 & 57963.00 & 0.59$^{+0.01}_{-0.01}$ & 1.88$^{+0.03}_{-0.03}$ & 74$^{+46}_{-19}$  & 0.50$^{+0.01}_{-0.01}$ & 15.2$^{+0.7}_{-0.7}$  & 1.11 & 0.38  & 0.73    & 1.12(3411) \tabularnewline 
{\it 24}$^{\ast}$ & 052 & 57963.09 & 57966.98 & 0.59$^{+0.01}_{-0.01}$ & 1.92$^{+0.02}_{-0.04}$ & >120              & 0.45$^{+0.01}_{-0.01}$ & 14.7$^{+0.8}_{-0.7}$  & 0.92 & 0.22  & 0.70    & 1.12(3411) \tabularnewline 
{\it 25}$^{\ast}$ & 056 & 57967.02 & 57977.99 & 0.59$^{+0.01}_{-0.01}$ & 1.80$^{+0.01}_{-0.03}$ & >150              & 0.21$^{+0.02}_{-0.01}$ & 42.7$^{+12.1}_{-10.2}$  & 0.60 & 0.08  & 0.52    & 1.12(3411) \tabularnewline 

{\it 26} & 057 & 57978.08 & 57982.99 & 0.51$^{+0.02}_{-0.02}$ & 1.74$^{+0.02}_{-0.03}$ & >122              & -                       &-  & 0.46 & -     &   0.46  & 0.95(278)  \tabularnewline 
{\it 27} & -   & 57983.06 & 58001.97 & -                      & 1.65$^{+0.05}_{-0.05}$ & >56               & -                       &-  & 0.24 & -     &   0.24  & 0.90(25)  \tabularnewline 
{\it 28} & 061 & 58002.03 & 58022.0 & 0.72$^{+0.08}_{-0.07}$ & 1.66$^{+0.05}_{-0.06}$ & >58               & -                        &- & 0.12 & -     &   0.12  & 1.08(71)  \tabularnewline 
{\it 29} & 066 & 58022.0  & 58075.38 & 0.70$^{+0.07}_{-0.07}$ & 1.68$^{+0.06}_{-0.06}$ &  35$^{+32}_{-10}$                & -        &-                 & 0.08 & -     &   0.08  & 1.23(101)  \tabularnewline 
                                                                  
\tabularnewline
\hline                                   
\end{tabular}}
\vspace*{3 cm}
\end{center}
\end{table}
\end{landscape}

\section{discussion}
\label{discussion}

\grs~ increases the sample of X-ray binaries that have shown ``failed" state transitions outburst, in which the source does not complete the full q-track pattern in the HID. In spite of the three softening events that occurred during the outburst, \grs~ did not make the transition to the canonical soft state (neither to the SIMS). This is supported by two findings: the disc luminosity never dominates the emission of the source and the fractional \rms~ never decreases below 10$\%$.

\cite{tetarenko16} reported on an all-sky database of galactic BHTs. They found that $\sim$40\% of sources are ``hard only", including also sources showing ``failed" state transition outbursts (i.e., HS-HIMS only). 
These authors suggest that the mass transfer rate ($\dot{\rm M}$) over the all outburst could be insufficient to allow the source to perform the transition to the soft states. Indeed, they found that all ``hard only" sources show outburst peak luminosities lower than 0.11\,L$_{\rm Edd}$.
The peak bolometric luminosity of \grs~ (L$_{\rm peak}\sim$0.03\,L$_{\rm Edd}$) is in agreement with their limit. 
On the other hand, the ``failed" state transition source \swf~\citep{delsanto16} showed L$_{\rm peak} \sim$0.37\,L$_{\rm Edd}$ \citep[d $\sim$7\,kpc;][]{munozdarias13}, which is more than a factor of three higher than the upper limit reported in \cite{tetarenko16}. 
This would suggest that the overall mass-transfer rate during the outburst is not the only parameter involved in the ``failed" state transition behavior or that the limit is higher than what \cite{tetarenko16} thought.\\ 
However, even though the most widely accepted distance of \swf~is 7\,kpc \citep{curran14, kalemci14}, this value is poorly constrained \citep{munozdarias13}.
Thus, if we assume a distance lower than $\sim$4\,kpc, even the BHT \swf~ would fall within the proposed upper limit of 0.11\,L$_{\rm Edd}$.

In the truncated disc model scenario, the inner radius of the accretion disc is smaller when we observe softer spectra.
In \grs\ this occurs up to a certain point when \kti\ starts to increase at constant inner disc radius.
Thus, we argue that \grs~ might have reached the ISCO in the HIMS.
We folded the XRT light curves at the binary system period ($\sim$14.7\,hr), covering about $\sim$80\% of this, and we did not find any evidence for dips and/or eclipses. In addition, we also folded the \integral/JEM-X light curve (3-6\,keV) collected during the 96\,ks of \integral\ ToO triggered by our team during a multi-wavelength campaign in the HS (results on this will be presented in a further paper). This light curve covered $\approx$100\% of the binary period and even in this case we did not find any evidence of features indicative of a high inclination system for \grs, which is likely to be ruled out.
Assuming an inclination angle the upper limit $\vartheta$ < 60$^\circ$ \citep{frank87} and using the inner disc radius R$_{\rm cos} \sim$15\,km, we obtain an upper limit R$_{\rm ISCO}$ < 21\,km. 
The lower limit on the BH mass \citep[M$_{BH}$ > 4.9\,M$_{\odot}$,][]{masetti96} allows us to estimate a lower limit on the gravitational radius as R$_{\rm g}$ > 7.3\,km which results in a R$_{\rm ISCO}$ < 3\,R$_{\rm g}$. In the Schwarzschild metric R$_{\rm ISCO}$ = 6\,R$_{\rm g}$, therefore the black hole in \grs~ would be rotating with a spin lower limit of $a_{*}>0.8$ (with $a_{*}$ the dimensionless spin).
We are aware that this is only a rough estimation, due to the extreme simplicity of the disc model used in this paper. 

The radio emission of \grs~ is consistent with coming from a compact jet, despite R$_{\rm in}$ being at the ISCO. 
However, we do not have an adequate radio coverage (e.g. a regular radio sample or VLBI observations) during the outburst and in particular during the soft episodes.
It is worth noting that, it does not appear that \grs~ got soft enough to quench the compact jet and to create transient ejecta \citep{fender04}.
The radio observations were performed when \grs~ was in HS-HIMS (see Sec. \ref{sec:radio}). We can not exclude that the jet changed its properties when the X-ray emission became softer and that there were rapid ejection events, quenching or steepening episodes we did not observe. This imply that the jet could be recovered in short times, between one radio observation and the next one.

An important tool to investigate the emission properties of BHTs is the radio/X-ray correlation. 
\cite{coriat11} suggested that the radio/X-ray correlation of the ``outliers" (or radio-quiet BHTs with L$_{\rm R}\propto$ L$_{\rm X}^{1.4}$) is produced by a radiatively efficient accretion flow. It means that L$_{\rm X}\propto\dot{\rm M}$, while the radio-loud branch (L$_{\rm R}\propto$ L$_{\rm X}^{0.6}$) would result from inefficient accretion where L$_{\rm X}\propto\dot{\rm M}^{2-3}$.
Both \grs~ and \swf~ are located on the steeper branch and they increase the number of radio-quiet BHTs.
Recently, the accretion efficiency picture proposed to explain the two different branches is challenged by the recent finding that  radio-quiet tend to have a negative spectral index $\alpha$ (S$_{\nu}\propto\nu^{\alpha}$) while for the radio-loud sources a positive $\alpha$ was observed. This suggests different core jet properties (rather than accretion flow) for the two classes of sources \citep{espinasse18}. 
Furthermore, we observe that the source never left the radio-quiet branch, again arguing against a quenching by the jet.\\
The power-law slopes of the \grs\ radio spectra, as also the \swf~ radio slope \citep[$\alpha\sim$ 0;][]{curran14}, are within the statistical distribution of the radio-quiet slope reported by \citet{espinasse18}. \\
Recently \cite{motta18} proposed that the origin of different radio loudness of BHTs is due to the geometric effects due to the inclination of the sources: i.e the radio-quiet sources would be at high inclination. However, \grs~ is a radio-quiet system most likely at low inclination. Nevertheless dips can be transient as observed in H 1743--322 (\citealt{motta15} and references therein), so the low inclination of \grs~ has to be confirmed. 
The radio-quiet/X-ray-bright behaviour of the source might be related to the very small inner disc radius that we infer from our disc modeling. Indeed, the small inner disc radius does not necessarily imply the disc extends uninterrupted from its outer parts and down to the ISCO. Coupled disc/corona condensation evaporation models, developed in the framework of the truncated disc model, predict that when the sources are close  to intermediate states, the inner hot-flow may re-condensate into an inner cool ring which would present observational signatures that are very similar to that of a full disc \citep{meyer09}. Then, the contribution from the soft photons of the disc would make the accretion flow brighter in X-rays with respect to an accretion flow at the same accretion rate (and therefore radio luminosity) in which this mini inner-disc would be absent. \cite{meyer14} argue that this could drive the different slope of the radio X-ray correlation of radio-quiet sources. Our results appear to be consistent with their picture. 

\section{Conclusions}
In this work we have presented the X-ray spectral and timing analysis of the BHT \grs~ during its 2016-2017 outburst. We analyzed the XRT and BAT observations collected during the whole outburst occurred from December 2016 to October 2017. In addition, we have reported on five radio observations and their correlation with X-ray data.
Our main results can be summarized as follows: 
\begin{itemize}
\item[1)] \grs~ can be added at the sample of the known BHTs that show a ``failed" state transition outburst.
During the outburst the source showed three softer episodes without making the transition to the soft state. 
Timing results and spectral parameters evolution are consistent with the source being in the HS at the beginning and at the end of the outburst, and in the HIMS during the spectral softening.
\item[2)] Our data suggest that the inner disc might have reached the ISCO during the three softening episodes, even-though the source was in the HIMS. 
However, disc/corona condensation-evaporation models predict that in the intermediate state the hot accretion flow may re-condensate into an inner mini-disc, mimicking the emission of the full disc.
\item[3)] The source might host a rapidly rotating BH with a rough spin lower limit of 0.8.
\item[4)] In the radio/X-ray luminosity plane \grs\ is located on the radio-quiet branch, which is in agreement with the X-ray brightness due to the condensed inner accretion flow. 
\end{itemize}

\section*{Acknowledgements}
We acknowledge financial contribution from the agreement ASI-INAF n.2017-14-H.0.
MDS acknowledges funding from the European Unions Horizon 2020 Programme under the AHEAD project (grant agreement nr. 654215).
TB thanks M. Capalbi for useful XRT data analysis support.
TB, MDS and AD thank A. Marino for the INTEGRAL/JEM-X data analysis.
TB and MDS thank M. Coriat for interesting scientific discussion on the X-ray/radio correlation.
This work made use of data supplied by the UK Swift Science Data Centre at the University of Leicester.
The Australia Telescope Compact Array and the Long Baseline Array are both part of the Australia Telescope National Facility which is funded by the Australian Government for operation as a National Facility managed by CSIRO. This work made use of the Swinburne University of Technology software correlator, developed as part of the Australian Major National Research Facilities Programme. This work was supported by resources provided by the Pawsey Supercomputing Centre with funding from the Australian Government and the Government of Western Australia.  The National Radio Astronomy Observatory is a facility of the National Science Foundation operated under cooperative agreement by Associated Universities, Inc.
JCAM-J is the recipient of an Australian Research Council Future Fellowship (FT140101082).
JM acknowledges support from PNHE in France, from the OCEVU Labex (ANR-11-LABX-0060) and the AMIDEX project (ANR-11-IDEX-0001-02) funded by the ``Investment d'Avenir" French government program managed by ANR.




\bibliographystyle{mnras}
\bibliography{biblio} 



\bsp	
\label{lastpage}
\end{document}